\documentstyle [12pt,epsfig]{article} 
\makeatletter
\@addtoreset{equation}{section}
\makeatother

\newcommand{\ba}{\begin{eqnarray}}
\newcommand{\ea}{\end{eqnarray}}

\begin{document}
\newcommand{\BS}{\bigskip}
\newcommand{\SECTION}[1]{\BS{\large\section{\bf #1}}}
\newcommand{\SUBSECTION}[1]{\BS{\large\subsection{\bf #1}}}
\newcommand{\SUBSUBSECTION}[1]{\BS{\large\subsubsection{\bf #1}}}

\begin{titlepage}
\begin{center}
\vspace*{2cm}
{\large \bf The Space-Time Lorentz Transformation:
Relativity of Simultaneity is Incompatible with Translational Invariance}  
\vspace*{1.5cm}
\end{center}
\begin{center}
{\bf J.H.Field }
\end{center}
\begin{center}
{ 
D\'{e}partement de Physique Nucl\'{e}aire et Corpusculaire
 Universit\'{e} de Gen\`{e}ve . 24, quai Ernest-Ansermet
 CH-1211 Gen\`{e}ve 4.
}
\newline
\newline
   E-mail: john.field@cern.ch
\end{center}
\vspace*{2cm}
\begin{abstract}
  Observations of the apparent times and positions of moving
  clocks as predicted by both `non-local' and `local' Lorentz
  Transformations are considered. Only local transformations
  respect translational invariance. Such transformations 
  change temporal but not spatial intervals, so breaking
  space time exchange symmetry and forbidding relativity
  of simultaneity and length contraction. A satellite
  cesium clock experiment to test these predictions is 
  proposed. 

\vspace*{1cm}
\end{abstract}
\end{titlepage}
 
\SECTION{\bf{Introduction}}
  The relativistic Length Contraction (LC)\footnote{In References~\cite{JHF1,JHF3}, by
  the present
  author, the acronym `LFC' for `Lorentz-Fitzgerald Contraction'was used. But
  as this name is more properly assigned to a conjectured dynamical effect in a 
  pre-relativistic
  theory (see Section 7 below), `LC' for Length Contraction or Lorentz Contraction
   (a consequence of the Lorentz Transformation) is used throughout the present paper.}
 and Time Dilatation (TD) effects were pointed out as consequences of the space-time 
  Lorentz Transformation (LT) in Einstein's original paper~\cite{Einstein1} on 
  Special Relativity (SR). The LC effect was clearly stated to be an `apparent'
  one in Reference~\cite{Einstein1}; even so, in many text books on SR it is
   stated to be not only an, in principle, experimentally observable effect but 
   also a `real' one implying that there is actually some dynamical contraction
   of the body, as discussed for example, in References~\cite{Sorensen,Bell}.
  \par It was only on making a more careful analysis of the physics of the
     observation process that it was realised, some five decades after Einstein's
   original paper, that the length contracted sphere that he considered would
   appear to be, to a distant observer, not flattened into an ellipsoid,
   but simply rotated~\cite{Terrell,Penrose}. This is because `observation'
   actually means the detection of photons emitted by, or scattered from, 
   the observed object. Not only the LC effect but also the the effects of
   light propagation time delays of the observed photons and optical
   aberration (that is, the change in the direction of motion of
   a photon due to the LT of its momentum) must also be properly accounted 
   for. In the present paper it is assumed throughout that all observations
   are corrected for the last two effects, so that the apparent positions
   are predicted by the LT only. 
   \par In a recent paper by the present author~\cite{JHF1} it was noticed that
    the LC effect itself is closely akin to the effect of light propagation
    time delays. In the latter, in order to arrive at the observer {\it at the same
    time} (this is the definition of the observation that gives LC) the photons
   coming from parts of the viewed object at different distances along the
   line of sight must be emitted from the object at different times. Similarly,
   because of the `relativity of simultaneity' proposed by Einstein, the photons
   recorded by the observer at a fixed time must be emitted, from the moving
   object, at different positions along the direction of motion of the 
   object. Thus LC is a strict consequence of the relativity of simultaneity.
   In the same paper~\cite{JHF1} two other apparent 
     distortions of space-time due to the LT: Space Dilatation (SD) in which
    the moving object appears longer (not shorter as in LC) and Time Contraction
    (TC) in which which the moving equivalent clock viewed at a fixed position
    in the observer's frame appears to be running faster (not slower as in TD)
    than an identical stationary clock, were pointed out. The four effects 
    LC, TD, SD and TC are all related to the projective geometry of the LT
    equations, and correspond to the projections with $\Delta t = 0$, 
    $\Delta x' = 0$,  $\Delta t' = 0$ and  $\Delta x = 0$ respectively
      \footnote{The space and time coordinates: $x$, $t$; $x'$, $t'$ are measured in two
     inertial frames, S; S' in relative motion along their common $x$-axis}.
    The `apparent' nature of all the effects is then manifest since a moving 
     object cannot be `really' contracted in, for example, the LC effect, if, with a 
     different observation procedure (SD) it appears to be elongated.
     A similar consideration applies to the (opposite) TD and TC effects.
     \par In an even more recent paper ~\cite{JHF3} the relation between the
     `real' positions of moving objects (i.e. those defined, for different
      objects in a common reference frame) to the apparent positions predicted
      by the LT was considered in detail. In particular, the well known 
      `Rockets-and-String'~\cite{DewBer} and `Pole-and-Barn'~\cite{Dewan} 
      paradoxes were re-discussed and a procedure for measuring the real 
     positions of moving objects in a single reference frame was proposed.
     \par The present paper pursues further the relation between the
      real and apparent positions of moving objects as well as analysing the
      times registered by two, spatially separated, synchronised, moving
      clocks viewed by a stationary observer. It is noted that, due to the
      ambiguity in the choice of the origin of spatial coordinates in 
      the rest frame of the moving objects or clocks,  no definite 
      predictions are obtained for either the times recorded by, or the
      apparent positions of, the moving clocks, but that, in every case,
      there is a violation of translational invariance. Only if a {\it local}
      LT is used, in which the origin of spatial coordinates in the rest frame
     of the moving object coincides with the spatial coordinate of the
     transformed event, is translational invariance found to be respected. 
     In this case there is no `relativity of simultaneity' and the related
     LC, TC and SD effects do not occur. However, TD, that results from a
     local LT, is unaffected by the restriction to this type of
     transformation.
      It is shown in Section 8 below that, in fact, TD is the only relativistic
       space-time effect that is confirmed experimentally. Only time intervals, 
       not space intervals, are modified by a local LT, thus breaking, in this case
       the space-time exchange symmetry recently proposed as a mathematical
       basis for SR~\cite{JHF2}. 
    \par The plan of this paper is as follows: In the following Section,
       translational invariance is discussed in a general way in connection
       with the observation of two identical and similarly accelerated clocks.
       In Section 3 the observation of the times of clocks subjected to a similar, 
       constant, acceleration in their proper frames is discussed for local
       and non-local LTs. In Section 4 the real and apparent positions of
       the moving clocks are discussed in relation to the LC effect. In
       Section 5, the principle of `Source Signal Contiguity' is proposed
       and some causal paradoxes of SR are resolved by use of the local LT.
       In Section 6 the breakdown of space-time exchange symmetry by a local
       LT is described and relativistic kinematics is discussed. Section 7
       contains an analysis of the Michelson Morley experiment considered 
       as a `photon clock'. A concise review of the experimental tests of
       SR performed to date is presented in Section 8. In Section 9 an
       extension of a previously performed satellite experiment is 
       proposed to measure the TC effect and therefore test, for the 
       first time, relativity of simultaneity.   

\SECTION{\bf{Moving Clocks and Translational Invariance}}

\begin{figure}[htbp]
\begin{center}\hspace*{-0.5cm}\mbox{
\epsfysize10.0cm\epsffile{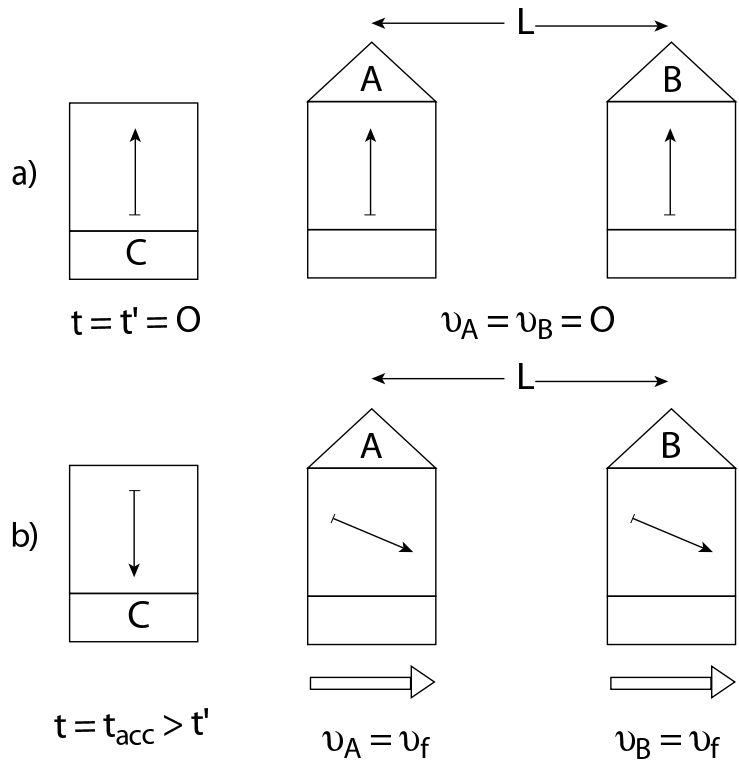}}
\caption{ {\em a) clocks A, B and C at $t = t'= 0$. They are at rest in S. b) After an identical
 acceleration program, during time $t_{acc}$ in S,  the clocks A and B (at rest in S') move
 in S with velocity $v_f$ parallel to the x-axis. Due to relativistic effects,
  $t' < t_{acc}$, but, from translational invariance, $t'_A  = t'_B$.}}
\label{fig-fig1}
\end{center}
\end{figure}
  
   In the following, in order to investigate the properties of the space-time LT, it will 
  be found convenient to consider two identical clocks, A and B, which perform identical
  motion parallel to the x-axis of a `stationary' inertial frame S. The common co-moving
  frame, at any instant, of A and B, is denoted by S'.
 Initially, A and B, which have each been
  synchronised with a reference clock, C, at rest in the frame S, are at rest in
   S, separated by a distance,
  $L$ (see Fig.1a). At time $t=0$ each clock is subjected to an identical acceleration 
   program during the time $t_{acc}$ in S. Because the velocities of A and B are the same
   at any instant, a common co-moving inertial frame always exists for the two clocks.
    Their separation, as measured 
     in this co-moving frame, or in the frame S, remains
    $L$ at all times\footnote{A method for performing such a measurement
    is described in Reference~\cite{JHF3}.}. Because of relativistic effects,
 that will be calculated in the
   following Section for a specific acceleration program,
 the observed time in S, $t'$, registered by
  the moving clocks A and B will differ from that shown by the reference clock C
  that is at rest in S. However, the times indicated by A and B must be identical.
  This is a consequence of translational invariance. The relativistic effects of the
  acceleration program must be the same whether they are applied to a clock initally at
  rest at position $x$ or one initially at rest at position $x+L$. After the time $t_{acc}$,
  the acceleration of each clock ceases and they continue to move with the same constant
  relativistic velocity $\beta_f$, separated by the distance $L$, as shown in Fig1b.

\SECTION{\bf{Time Dilatation of Moving Clocks using `Local' and `Non-Local'
          Lorentz Transformations}}

\begin{figure}[htbp]
\begin{center}\hspace*{-0.5cm}\mbox{
\epsfysize10.0cm\epsffile{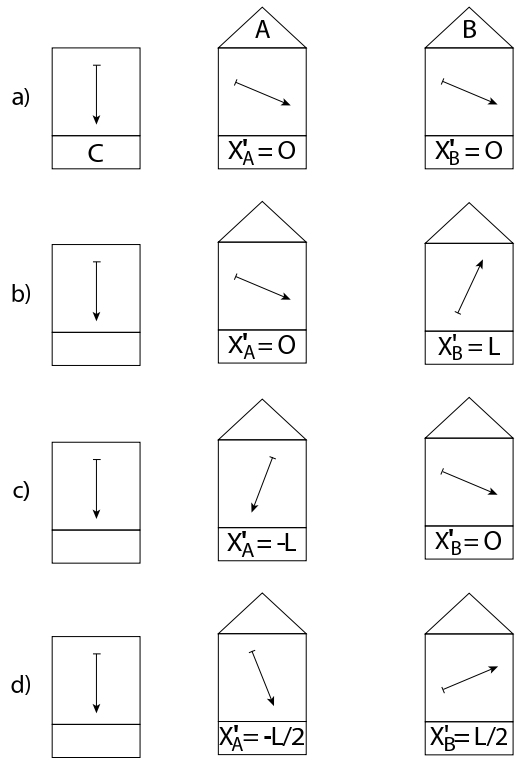}}
\caption{{\em Times indicated at time $t = t_{acc}$ in S
  by the clocks A, B and C, according to the space-time
  LT, for different choices of the origin, $O'$, of S': a) local LT for both A and B
 ($x'_A = x'_B = 0$),   b)  $O'$ at A,  c)  $O'$ at B,  d)  $O'$ midway
  between A and B. Only a) gives a prediction consistent with 
  translational invariance. Units with $a = c = 1$ are used and $v_f = \sqrt{3}/2$,
  $\gamma_f = 2$,
  $t_{acc} = \sqrt{3}$.}}
\label{fig-fig2}
\end{center}
\end{figure}

 It is now assumed that the clocks A and B, introduced above, initially at rest in S,
  are subjected to the
 same uniform acceleration, $a$, in their own rest frames, at the same instant in S, in
 the direction of the positive x-axis, for a time, in S, of duration $t_{acc}$. The
  formulae giving the relativistic velocity, $\beta(t) = v(t)/c$, and position, $x(t)$,
  in the frame S as a function of the elapsed time, $t$, in this frame have been give by,
  for example, Marder~\cite{Marder} and Nikolic~\cite{Nikolic}:
  \begin{eqnarray}
    \beta(t) & = & \frac{at}{c\sqrt{c^2+a^2t^2}}  \\
     x(t) &  = & \frac{c}{a}\left[\sqrt{c^2+a^2t^2}-c\right]
  \end{eqnarray}
 The observed time, $t'$, indicated by the clocks in S , as a function of $t$
 is calculated by integrating the differential form of the LT of time in
 accordance with the time variation of $\beta$ given by Eqn(3.1). In the first case
 a {\it local} LT is used for each clock. This corresponds to the choice
 $x' = 0$ in the general space-time LT:
\begin{eqnarray}
x' & = & \gamma (x-vt) \\
t' & = & \gamma (t-\beta \frac{x}{c}) \\
x & = & \gamma (x'+vt') \\
t & = & \gamma (t'+\beta \frac{x'}{c})
\end{eqnarray}
 That is, the origin of coordinates in S' is chosen to be at the position
 of the transformed space point (the position of the clock A or B). A {\it non-local}
 LT is one in which the origin of coordinates in S' is not at the same position
 as the transformed space point. There are clearly an infinite number of such LT
 equations for any transformed space point, corresponding to an arbitary choice of 
 origin in S'.
 Situating clock A at $x'=0$,
 the differential form of (3.6) corresponding to Eqn(3.1) is:
 \begin{equation}
 dt' = \left[\frac{1}{\gamma(t)}-\frac{t}{\gamma(t)^2}\frac{d \gamma(t)}{dt}\right] dt
 \end{equation}
 where:
  \[ \gamma(t) = \frac{1}{\sqrt{1-\beta(t)^2}} \]
 or, using Eqn(3.1),
 \begin{equation}
 dt' = \left[\frac{1}{\sqrt{1+\frac{a^2t^2}{c^2}}}-\frac{a^2 t^2}
{c^2\left(1+\frac{a^2t^2}{c^2}\right)^{\frac{5}{2}}}\right] dt
 \end{equation} 
 Performing the integral over t gives:
 \begin{equation}
 t'(t, x'=0) = \frac{c}{a} \left[ \ln \left(\frac{at}{c}+\sqrt{1+\frac{a^2t^2}{c^2}}
\right) - \frac{a^3 t^3}
{3c^3\left(1+\frac{a^2t^2}{c^2}\right)^{\frac{3}{2}}}\right]
 \end{equation} 
\[ {\rm local~LT}, ~~x' = 0,~~ 0 < t < t_{acc} \]
 For $t \ge t_{acc}$ Eqn(3.7) simplifies to:
  \begin{equation}
 dt' = \frac{dt} {\gamma(t_{acc})}
 \end{equation}
 so that for $ t \ge t_{acc}$,
  \begin{equation}
  t' = t'(t_{acc}, x'=0) + \frac{t- t_{acc}}{\gamma(t_{acc})} 
 \end{equation}
 where $ t' = t'(t_{acc}, x'=0)$ is given by Eqn(3.9).
 An identical result is, of course, given by a local LT for clock B.
 Thus, in this case, translational invariance, as depicted in Fig.1a, is evidently
 respected. 
 \par A non-local LT is now used to evaluate $t'$ for the clock B. The origin 
 of the LT in S' is chosen at the position of the clock A. The value of
 $t'$ for A, $t'_A$, is
 then given by Eqns(3.9),(3.11). Now the position of B correponds to a non-local
 LT with $x'_B = L$. In this case, Eqn(3.8) is replaced by the expression:
 \begin{equation}
 dt' = \left[\frac{1}{\sqrt{1+\frac{a^2t^2}{c^2}}}-\frac{a^2 t^2}
{c^2\left(1+\frac{a^2t^2}{c^2}\right)^{\frac{5}{2}}}\right] dt -\frac{L}{c} d \beta
 \end{equation}   
 and, on integrating over $t$:
 \begin{equation}
 t'_B(t, x_B'= L) = \frac{c}{a} \left[ \ln \left(\frac{at}{c}+\sqrt{1+\frac{a^2t^2}{c^2}}
\right) - \frac{a^3 t^3}
{3c^3\left(1+\frac{a^2t^2}{c^2}\right)^{\frac{3}{2}}}\right] - \frac{L a t}
{c^2\sqrt{1+\frac{a^2t^2}{c^2}}} 
\end{equation}   
  \[ {\rm non-local~LT}, ~~x'_B = L,~~ 0 < t < t_{acc} \] 
 That is:
 \begin{eqnarray}
 t'_B(t, x_B'= L) &  =  & t'_B(t, x_B'= 0) - \frac{L a t}
{c^2\sqrt{1+\frac{a^2t^2}{c^2}}} \nonumber \\
  & = &  t'_B(t, x_B'= 0) - \frac{L \beta(t)}{c} 
\end{eqnarray}   
 Since, for the local LT with $x' =0$ for both A and B:
 \begin{equation}
 t'_A(t, x_A'= 0) =  t'_B(t, x_B'= 0) 
\end{equation}                
 it follows that:
 \begin{equation}
 t'_B(t, x_B'= L)   =   t'_A(t, x_A'= 0) - \frac{L \beta(t)}{c}
\end{equation}
  Evidently $t'_A$ and $t'_B$ are different, and so translational invariance
 is not respected if $L \ne 0$, i.e. if the LT applied to the clock B is non-local.
 \par Alternatively, choosing the origin in S' of the LT for clock A to coincide
  with the position of B ($x'_A = -L$) or to lie midway between the two clocks
  ($x'_A = -L/2$, $x'_B = L/2$) gives the results, valid in the interval
 $t \le t_{acc}$:
 \begin{eqnarray}
 t'_A(t, x_A'= -L) &  =  & t'_A(t, x_A'= 0) +\frac{L \beta(t)}{c} \\
 t'_A(t, x_A'= -L/2) &  =  & t'_A(t, x_A'= 0) +\frac{L \beta(t)}{2c} \\ 
 t'_B(t, x_B'= L/2) &  =  & t'_A(t, x_A'= 0) -\frac{L \beta(t)}{2c} 
\end{eqnarray}
 The times registered by the clocks A and B, as viewed from S, at time 
  $ t =  t_{acc}$, are shown, for the different choices of the origin in S'
   of the
  LT considered above, in Fig.2. Units are chosen such that $c = a = 1$. Also
  $L =1$ and $\beta_f = \beta(t_{acc}) =  \sqrt{3}/2$, so that
  $\gamma_f = 1/\sqrt{1-\beta_f^2} = 2$ and $t_{acc} = \sqrt{3}$ .
 The time interval
  $L\beta_f/c = \sqrt{3}/2$ corresponds to a $90^{\circ}$ rotation of the hands 
  of the clocks A and B, while $t_{acc} = \sqrt{3}$ corresponds to a 
   $180^{\circ}$ rotation of the hand of C.
  \par Since the procedure described for accelerating the clocks is
  physically well defined and unique, the times indicated by the clocks
  A and B must also be unique, so at most one of the four cases shown in
  Fig.2 can be physically correct. Since it has been argued in Section 2
  above that translational invariance requires that the clocks A and B
  indicate always the {\it same} time, the only possiblity is that shown in 
  Fig.2a, corresponding to a local LT for each clock. The non-local
  LT used in Figs.3b,c,d must then be unphysical. In each of these
  cases where events contiguous in space-time with one clock, or both
   clocks, are subjected to a non-local
  LT, it can be seen that clock A is apparently in advance of 
  clock B by the fixed time interval  $L\beta_f/c$. This is
  just the apparent effect of `relativity of simultaneity' 
   resulting from the different spatial positions of the
  clocks introduced in Einstein's first paper on SR~\cite{Einstein1}.
  It results from the term $\beta x'/c$ in Eqn(3.6).
  The argument just presented shows that the physical existence of
  such an effect is in contradiction with translational invariance.
  It must then be concluded that, if translational invariance
  is respected, a non-local space-time LT is 
  unphysical and that `relativity of simultaneity' does not exist.
  Since LC is a direct
  consequence of the relativity of simultaneity~\cite{JHF3}
  it also cannot exist if translational invariance is
  respected. The LC effect is further discussed in the following
  Section.
  \par It may be noted that in all cases shown in Fig.2, the formula
   (3.11) is valid, on substituting the appropriate value of $t'(t_{acc})$.
    If $t_1 > t_2 > t_{acc}$ it then follows, independently of the 
    choice of a local or non-local LT, or of the position of the
    origin of the latter in S', that: 
 \begin{equation}
 \Delta t' =  t_1'-t_2' = \frac{ t_1-t_2}{\gamma(t_{acc})} = 
 \frac{\Delta t}{\gamma(t_{acc})}
\end{equation}
  Thus the rate, as observed in S, of both clocks A and B, after acceleration,
 is slowed
 down in accordance with the well known Time Dilatation (TD) effect, although
  the actual times recorded by the clocks do depend on the choice of the
  S' origin for the non-local LT. Indeed, as shown in Fig.2c, for the choice 
  $x'_A=-L$, the clock A is apparently {\it in advance} of the stationary
  clock C at the end of the acceleration period. The TD effect for
  uniformly moving clocks is given by the $\Delta x' = 0$ projection
  of the LT~\cite{JHF1} and, as shown in Section 4 below, is defined
   as the result of
  successive {\it local} LTs performed on the moving clock.
  Thus the TD effect is perfectly compatible with translational
  invariance.
  \par It is important for the calculation of $t'(t)$ that the clocks
  A, B, C be properly synchronised at time $t =0$ when they are at rest
  (see Fig.1a). As discussed at length in Reference~\cite{Einstein1} the 
  problem of synchronisation of clocks, situated at different spatial
  positions, is a non-trivial one. In fact in Reference~\cite{Einstein1}
  the `relativity of simultaneity' was derived by comparing the
  synchronisation of clocks in a stationary and a moving frame. However,
  it is not necessary to discuss clock synchronisation to understand
  that the physical situation depicted in Fig2a is the only possible
  one. For this it is convenient to introduce `radioactive clocks'
  which need no synchronisation. If the clocks A, B and C consist
 of similar samples of radioactive nuclei, equipped with a detector,
  then the proper time $t'$ is determined by the number of radioactive
  decays, $N_{obs}$, recorded. In the limit that $N_{obs} \rightarrow \infty$:
 \begin{equation}
  t' = \tau_N \ln\left(\frac{N_0}{N_0-\epsilon N_{obs}}\right) 
\end{equation}
 where $N_0$ is the number of unstable nuclei at time $t = 0$, $\tau_N$
 is the mean life of the radioactive nucleus and $\epsilon$ is the
 efficiency of detection. Allowing for the statistical accuracy in
 the determination of $t'$ due to the finite value of $N_{obs}$,
  it is clear that two such clocks, initially situated at different
  spatial positions and then subjected to identical
  acceleration programs, must record equal values of $t'$ at 
  at any time $t$ in S. Thus Figs.2b,c,d clearly correspond to physically
  impossible situations.
 
\SECTION{\bf{Local and Non-Local Lorentz Transformations and the
  Relativistic Length Contraction}}

\begin{figure}[htbp]
\begin{center}\hspace*{-0.5cm}\mbox{
\epsfysize10.0cm\epsffile{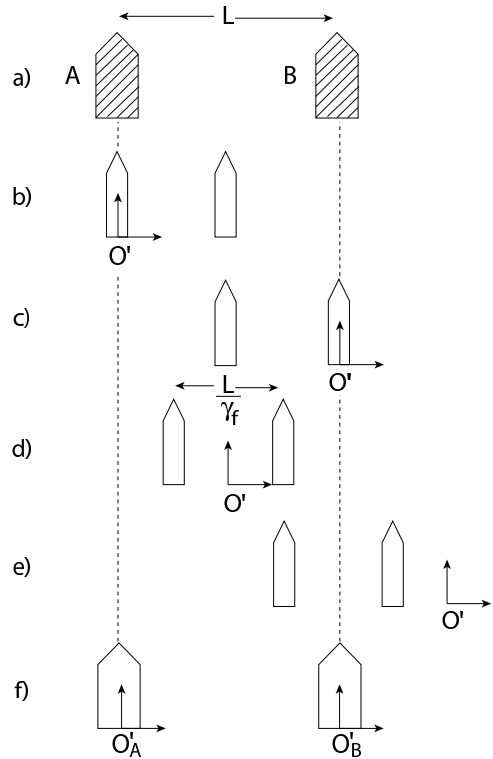}}
\caption{ {\em a) real positions of clocks A and B in either S or S'. b)-f) show the
  apparent positions in S of the clocks A and B for different choices of the origin,
 $O'$, in S': b) $O'$ at A, c) $O'$ at B, d) $O'$ midway
  between A and B. A, e) $x'_A = -3L/2$, f) local LT for both A and B
  ($x'_A = x'_B = 0$). Units and parameters as in Fig.2}}
\label{fig-fig3}
\end{center}
\end{figure}

 In order to discuss the LC effect it will be 
 convenient to consider the positions of the clocks A and B,
 introduced above,  for $t > t_{acc}$. Both
 clocks then move with their final constant relativistic velocity 
 $\beta_f = \beta(t_{acc})$. The separation of the clocks, in both S and S',
 is $L$, as a consequence of the identical acceleration program to which 
  they were subjected. As in the case of the calculation of $t'(t)$
 discussed in the previous Section, different LT, both local
  and non-local, will be used to calculate the apparent 
  positions of the clocks in S and the clock synchronisation between
  S and S' will be performed during the phase of uniform motion following
  acceleration. 
  \par The first case considered is same as an early discussion of the LC effect
  by Einstein~\cite{Einstein2}, and is similar to the presentation of 
  LC given in most text books on SR. The origin of the coordinate system in
  S is chosen to coincide, at some instant when the clocks are synchronised
  so that $t = t' =0$, with that of S', which is located at the position
   of clock A. Applying the LT Eqn(3.3) then gives $x_A^{app} = 0$
   at this time, i.e. the 
   apparent position, according to the local LT at the clock A is the same
   as the real position of clock A at $t=0$. The LT (3.3), with the above
   choice of clock synchronisation, is now applied to the clock B. This 
  non-local LT with $x' =L$ gives, for the apparent position of B,
  $x_B^{app} = L/\gamma_f$. Thus, the apparent position of B, according to the LT,
  is shifted from the actual position, $x_B = L$  of B at $t = 0$.
  The distance between the apparent positions of A and B is:
  $x_B^{app}-x_A^{app} = L/\gamma_f-0 = L/\gamma_f$, the well-known LC effect. The results
  of repeating this type of calculation, with different choices of the origin
  in S', but always synchronising the clocks so that $t = t' =0 $ when the origins
  of S and S' 
  coincide, are presented in Table 1 and shown in Fig.3. 
  Fig.3a shows the real positions of the clocks in either S or S'. In 
  Fig3b-f the apparent positions of the clocks according to the LT, as observed
  in S, are shown for different choices of the origin O' of the LT.
  In Fig.3b, corresponding to Einstein's calculation of Reference~\cite{Einstein2}
  the LT is local for A but not for B, In Fig.3c it is local for B but
  non-local for A. In Figs3d,e it is non-local for both clocks. Finally, in Fig.3f
  it is local for both clocks. Since, as shown in the previous Section, only the
  case shown in Fig.3f is consistent with translational invariance of $t'(t)$
  this is the only physically possible solution. In can be seen that, in this case,
  the real and apparent positions of the clocks are the same and there is no LC effect.
  In fact, it is assumed that the local LT are performed for {\it all points}
  of the spatially extended clocks A and B. Thus the apparent sizes of the
  clocks are the same in S and S' (see Fig.3f). If a fixed origin is
  chosen in S' for the LT of events contiguous in S' with 
  the clocks, as in Figs.3b-e, the clocks are apparently
  contracted, parallel to their direction of motion, by the factor
  $1/\gamma_f$.

  \par Indeed it is clear by an even more basic requirement of a physical theory,
   that it gives {\it some} well defined prediction, that the LC cannot be even 
   an apparent physical phenomenon. There are an infinite number of different
   predictions for the apparent positions of the moving clocks, in all of which
   they are separated by the `Lorentz-contracted' distance $L/\gamma_f$, differing
   only by the arbitary choice of the position of the origin, O', of the non-local
   LT. In fact the apparent position of, say, $x_A$, can be {\it anywhere} on the x-axis
   with a suitable choice of O'. As can be seen from the fourth row of Table 1, 
   $x_A = X_A$ and $x_B = X_A+L/\gamma_f$, for any $X_A$, by choosing
    $x'_A = -\gamma_f X_A/(\gamma_f-1)$. The problem here can be formulated in terms
    of a general principle that should be obeyed by any acceptable physical theory.
    It might be called CIPP for `Coordinate Independence of Physical Predictions':
    \par{\tt The predictions of physical phenomena by any theory must be independent
         of any arbitary choice of coordinate system.} 
     \par It is shown clearly in Fig.3 that this principle is not respected by a non-local LT.
     Since, for a local LT the arbitariness in the choice of the origin of 
     coordinates is, by definition, absent, CIPP is not applicable and unique predictions
     for the observed of space-time positions of
     events in S are always obtained. 
\begin{table}
\begin{center}
\begin{tabular}{|c||c|c|c|c|c|c|} \hline
  $x'_A$  & $x_A^{app} - x_A$ &  $x_B^{app} - x_A$ & $t_A$ & $t_B$ &
  $t'_A$ & $t'_B$  \\ \hline \hline
 0 & 0 & $\frac{L}{\gamma_f}$ & 0 & 0 & 0 & -$\frac{\beta_f L}{c}$ \\
   &  &  &  &  &  &   \\
 -$L$ & $\frac{(\gamma_f-1)L}{\gamma_f}$ & $L$  & 0 & 0  & $\frac{\beta_f L}{c}$ & 0 \\
   &  &  &  &  &  &   \\
 -$\frac{L}{2}$ & $\frac{(\gamma_f-1)L}{2\gamma_f}$ & $\frac{(\gamma_f+1)L}{2\gamma_f}$  &
   0 & 0  & $\frac{\beta_f L}{2c}$ & -$\frac{\beta_f L}{2c}$  \\
   &  &  &  &  &  &   \\
 -$(L+D)$ & $\frac{(\gamma_f-1)(L+D)}{\gamma_f}$ & $L+ \frac{(\gamma_f-1)D}{\gamma_f}$  &
   0 & 0  & $\frac{\beta_f(L+D)}{c}$ & $\frac{\beta_f D}{c}$  \\ 
   &  &  &  &  &  &   \\         
\hline
\end{tabular} 
\caption[]{{\em Apparent positions and times of clocks A and B for different 
 choices of origin in S'. In all cases the clocks in S and S'
  are synchronised at $t = t' = 0$ when the origins of S and S' are
  coincident.}}     
\end{center}
\end{table}
 \par As will be discussed in Section 8 below, and unlike for TD, there is no
  experimental evidence for the relativity of simultaneity or the LC. The same is true
  of the two other effects of SR~\cite{JHF1}, Space Dilatation (SD), the $\Delta t' = 0$ 
 projection of the LT, and Time Contraction (TC) the $\Delta x = 0$ projection
  of the LT, both of which are also direct consequences of the relativity of simultaneity.
  On the other hand, the well verified TD effect is, by definition, the prediction
  of a {\it local} LT. Typically, 
  the moving clock is assumed to be situated at $x' = 0$ so that the LT of Eqns(3.3) and
  (3.4) becomes:
 \begin{eqnarray}
x' & = & 0 \\
t' & = & \frac{t}{\gamma} \\
x & = & \gamma vt' \\
t & = & \gamma t'
\end{eqnarray}  
 \[ {\rm Local~LT~of~an~event~at~}~x' = 0 \]
 Eqns(4.2) and (4.4) are identical and , by use of Eqn(4.2),
 Eqn(4.3) reduces to:
\begin{equation}
  x = vt      
\end{equation}
which is just the trajectory of the origin
 of  S' in S. Thus, unlike in the case of a non-local LT,
 there is no distinction between the apparent position of the
 observed event in S and the actual position, in this frame, of
 the corresponding point in S'.
 Eqn(4.4) is just the TD effect.
 \newpage
 \par The inverse local LT is:
 \begin{eqnarray}
x & = & 0 \\
t & = & \frac{t'}{\gamma} \\
x' & = & -\gamma vt \\
t' & = & \gamma t
\end{eqnarray}
 \[ {\rm Local~LT~of~an~event~at~}~x = 0 \]   
 Again, Eqns(4.7) and (4.9) are identical, and Eqns(4.7) and
 (4.8) may be combined to give:
\begin{equation}
  x' = -vt'      
\end{equation}
 the trajectory of the origin of  S in S'. Eqn(4.9)
 is the TD effect for a clock at rest in S when viewed 
  from S'.
 \par Since the spatial position of the transformed event
 can always be chosen at the origin of S', it is
 proposed here that Eqns(4.1)-(4.10) embody the essential physical content
  of the 
 space-time LT. Thus, only time and not space is modified
 in the passage from Galilean to Special Relativity.
 The local LT differs from the general LT of Eqns(3.4) and (3.5)
 only in the specific choice of origin for the transformed
 event. As seen above, SR based on the local LT (4.1)-(4.4)
 gives, unlike the non-local LT, a unique prediction that respects
 translational invariance. In such a theory there is no relativity
 of simultaneity or the associated apparent distortions of space
 and time: LC,~SD,~TC. The apparent lengths of physical objects, or their 
 spatial separations, are the same in all inertial frames, and
 correspond to the real positions of the objects, described by
 the usual Galilean kinematical laws, in each frame. Only the
 apparent time, described by the TD formula, changes from 
 one inertial frame to another.
 In the following Sections the differences in space-time geometry
 of the local and non-local LT are explored in more detail, as well
 as relativistic kinematics, which is shown to be unchanged 
 by the restriction to a local LT. 

\SECTION{\bf{`Source Signal Contiguity' and the Resolution
 of Some Causal Paradoxes of Special Relativity}}

 In Section 2 above, the breakdown of translational invariance by a
 non-local LT was demonstrated by considering the apparent times of
 similarly accelerated clocks. This breakdown is however directly evident, in
 space, in Figs3a-3e. It is sufficient to consider Fig3b, corresponding
 to Einstein's demonstration~\cite{Einstein2} of the LC. The apparent position
 of A is the same as the real position of the clock, whereas for the indentical
  clock B, only displaced by the distance $L$ along the x-axis, the apparent
  position differs by $L/2$ from its real one. Since the clocks are identical
  apart from their position along the x-axis, the relationship between the 
  real and apparent positions must be the same for both clocks. Since the non-local
  LT predicts that this is not the case it clearly violates translational
  invariance. 
  \par Consider now single photons emitted (or reflected) from A and B, which, when
   detected in S, constitute an observation of the apparent positions of the 
   clocks in this frame. In S', the proper frame of the clocks, it is always
   assumed that the space-time events corresponding to the photon emission
   process and that describing the position of the source, are contiguous.
   The same is true, in the example of Fig3b, for the space time event 
   corresponding to the observation of the photon from A in S, and that 
   corresponding to the real position of A at the time of observation.
  This is not the case for the
   photon emitted by B. Not only is it spacially displaced from the real position
   of B at the time of observation, but also, as shown in the 
  first row of Table 1, it is emitted from B at the earlier time, 
   $t' = -\beta_f L/c$, when the real position of B is displaced in S by the distance
   $-\gamma_f \beta_f^2 L = -3/2$, as compared to
   that shown in  Fig3a, where the clocks A and B 
   are separated by one spatial unit. Thus the photon appears to be `hanging in 
    time' in S for the period $\gamma_f \beta_f L/c = \sqrt{3}$ between its times of 
   emission and observation.
   \par A similar consideration of the example shown in Fig.3d, presented elsewhere
   ~\cite{JHF3}, shows that the photon emitted from A is apparently observed
    in S {\it before} A reaches the position at which the photon is emitted
    in S', thus apparently violating causality. None of these paradoxes occur 
    when local LT at A and B are used to calculate the apparent positions
    of the photons in S as shown in Fig.3f. The observed positions of the
     photons are then contiguous in space-time with  the real positions of
     A and B. 
    \par The causal paradox of the `backward running clocks' pointed out in a recent
     paper~\cite{Soni} is also resolved by use of the local LT. In Fig.1c of 
     Reference~\cite{Soni}, the different apparent times registered by the clocks
     A',B' and C' are due to the term $\beta x'/c$ in Eqn(3.6) that leads to
     `relativity of simultaneity'. This figure shows a similiar effect to that
      of Fig2b,c,d of the present paper. In order that the three clocks
      show the {\it same} time when they are brought to rest, as shown in
      Fig.1d of Reference~\cite{Soni} (thus, tacitly, imposing the condition 
      of translational invariance) it is necessary that the
      clock A' apparently runs `backward in time' and C' `forward in time' 
      during the deceleration. If local LT are used to calculate the
       apparent times of A' and C' (as is already done, in the example, for B')
        all the moving clocks in  Fig.1c of Reference~\cite{Soni} indicate 
        the same apparent time and the paradox is resolved.
     \par Since essentially all our knowledge of objects in the external world
     is derived from our observation (direct or indirect) of photons emitted
      by, or scattered from, them, it is only possible to obtain knowledge of the
      real positions or sizes of such physical objects by making an assumption,
      that may be called `Source Signal Contiguity' (SSC). The latter may
      be stated as the following physical principle:
      \par {\tt Space-time events corresponding to photon emission processes
          and the \newline space-time positions of their sources are contiguous in
          all frames of \newline observation.}
       \par This is the tacit assumption made, for example, in all astronomical
      measurements where, say, the diameter of the sun or moon are deduced from
      a pattern of photons detected by a telescope. As shown above, the SSC
      principle is not respected by the non-local LT. To give another 
      concrete example of this in Astrophysics, consider a star moving perpendicular
      to the line of sight
      with velocity $\beta c$ relative to the earth, that explodes,
      emitting a light pulse of very short duration in its proper
      frame. As viewed from the earth,
      with coarse time resolution, the image of the explosion will be, according to
      a calculation using a non-local LT,   
      elongated parallel to the velocity direction by the factor $\gamma$
      due to the SD effect~\cite{JHF1}. In S, 
     the observed photons from the star would not respect the SSC principle.
      A calculation using a local LT for each emitted photon, after applying
      appropriate
     corrections for light propagation time delays and optical aberration,
     does respect SSC and, as in Fig3f, predicts an observed image that
     faithfully reflects the real spatial distribution of the different
     photon sources in the frame of observation. 

\SECTION{\bf{Local Lorentz Transformations and Relativistic Kinematics}}
\begin{figure}[htbp]
\begin{center}\hspace*{-0.5cm}\mbox{
\epsfysize10.0cm\epsffile{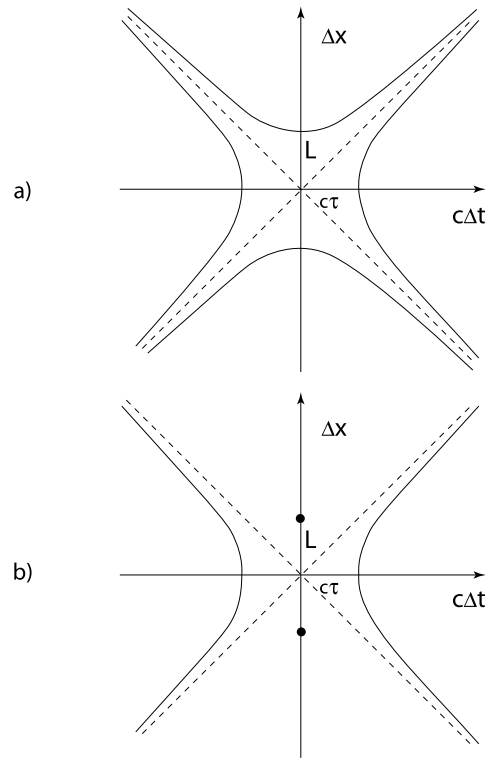}}
\caption{{\em a) Minkowski plot for non-local Lorentz Transformations. There is symmetry
   for $\Delta x \leftrightarrow c \Delta t$, and for space-like separations
   $\Delta t$ may be positive or negative. b)  Minkowski plot for local Lorentz
   Transformations. For space-like separations $\Delta t = 0$, and the symmetry under 
    $\Delta x \leftrightarrow c \Delta t$ is broken.}}
\label{fig-fig4}
\end{center}
\end{figure} 
   It may be remarked that the procedure of deriving a local LT from a non-local
 one is analagous to the clock synchronisation procedure used to derive the 
 standard space-time LT from the more general one in which the origins,$O$ and $O'$, of
 the space-time coordinate axes are explicitly given:
  \begin{eqnarray}
  x' - x'(O') & = & \gamma \left(x-x(O)-v(t-t(O) \right) \\
  t' - t'(O') & = & \gamma \left(t-t(O)-\frac{\beta(x-x(O))}{c} \right)
 \end{eqnarray}   
  Fixing first the origin of S at $x(O) = t(O) = 0$ gives:
   \begin{eqnarray}
  x' - x'(O') & = & \gamma(x-vt) \\
  t' - t'(O') & = & \gamma(t-\frac{\beta x)}{c})
 \end{eqnarray}     
  so that:
   \begin{eqnarray}
  x'(x=0,t=0) & = & x'(O') \\
  t'(x=0,t=0) & = & t'(O')
 \end{eqnarray}    
  The conventional choice, yielding Eqns(3.3) and (3.4) is 
   $ x'(O') = t'(O') = 0$. The local LT of Eqns(4.1) to (4.5)
   is given, instead, by the choice: $ x'(O') = x', t'(O') = 0$.
  \par  The LT expresses a certain symmetry in space-time which is
   most evident in the Minkowski~\cite{Minkowski} plot of 
   $\Delta x = x_1-x_2$, versus $ c\Delta t = c(t_1-t_2)$, which shows the
   loci of $\Delta x$ and $\Delta t$ as observed from different inertial
   frames. Every pair 1,2 of events in space-time is either
   space-like separated (if a LT exists such that $\Delta t = 0$) 
   or time-like separated (if a LT exists such that $\Delta x = 0$).
   The Minkowski plot corresponding to a non-local LT, i.e. where
   a common origin in S' is assumed for each pair of spatially separated events, 
   is shown in Fig.4a. Each pair of space-time points defines defines a
    hyperbola in $\Delta x$, $c\Delta t$ space. The hyperbolae of space-like
    (time-like) separated pairs intersect the $\Delta x$ (c$\Delta t$) axes 
    at right angles. Einstein's `relativity of simultaneity' is reflected
  in the different possible signs of $\Delta t$ for space-like separated
   pairs. The local LT gives the Minkowski plot shown in Fig4b.
   The hyperbolae describing points with time-like separation are unchanged,
    whereas those for space-like separations condense to two points 
    on the $\Delta x$ axis. For time-like separated points, the equations
    of the hyperbolae are: 
  \begin{equation}
  c^2\tau^2 = (c\Delta t)^2 - (\Delta x)^2
  \end{equation}
 Using Eqn(4.5),
  \begin{equation}
   \Delta x = v \Delta t
  \end{equation}
 so that,
  \begin{equation}
  \tau^2 = (\Delta t)^2(1-\frac{v^2}{c^2}) = \frac{(\Delta t)^2}{\gamma^2}
    = (\Delta t')^2 
  \end{equation}   
  where Eqn(4.4) has been used, and S' is the frame where $\Delta x = 0$. The equations
 of the hyperbolae for space-like separated points are: 
  \begin{equation}
  L^2 = (\Delta x)^2 - c^2 (\Delta t)^2 
  \end{equation}
 Applying a local LT to the points 1,2 in the frame S', where $t'_1 = t'_2$ gives,
 from Eqn(4.4):
   \begin{eqnarray}
 t_1 & = & \gamma t'_1 \\
 t_2 & = & \gamma t'_2
 \end{eqnarray} 
or since $t'_1 = t'_2$,
  \begin{equation}
  \Delta t = t_1 -t_2 = \gamma (t'_1- t'_2) = 0 
  \end{equation}  
Thus for all values of $\gamma$, i.e. in every inertial frame,
 $\Delta t = 0$, so that the hyperbolae In Eqn(6.10) reduce to: 
  \begin{equation}
 \Delta x = \pm L  
  \end{equation}
  and the spatial separation of the events is observed to be
 the same in all inertial frames.  
 Comparison of Fig.4a with 4b, shows that the restriction to a local LT
  imposed by translational invariance has the effect of breaking
  the symmetry of SR under the exchange of space and time~\cite{JHF2}. 
  The form of Fig.4a remains unchanged under the exchange:
 $\Delta x \leftrightarrow c\Delta t$, but this is no longer the case in Fig.4b, 
  where the hyperbolae crossing the $\Delta x$ axis are reduced to points at 
  $\Delta t = 0$.
 \par In order to introduce the description of kinematics by SR, the two 
 arbitary space-time events 1, 2 just discussed may be replaced
  by similar space-time events lying along
  the space-time trajectory (world-line) of a massive physical object,
   O($m$), of Newtonian mass $m$, at rest in S'. Such events are time-like
   separated and so lie on the hyperbolae intersecting the $c \Delta t$ axis 
   at right angles in both Fig.4a and Fig.4b.
 If the object 
  is situated at $x' = 0$ in S', 
   then $\Delta x' = x'_1-x'_2 = 0$ and the local LT (4.4) gives:
\begin{equation}
\Delta t =  t_1-t_2 = \gamma(t'_1-t'_2) = \gamma \Delta \tau
\end{equation}
 where $\tau \equiv t'$ is the proper time of the object.
 The energy momentum and velocity 4-vectors, $P$ and $V$ respectively of O($m$)
 are defined as~\cite{Einstein3}: 
 \begin{eqnarray}
 P & \equiv & m \frac{dx}{d \tau} = (P_0,P_x) = mV  \\
 V & \equiv & (c \gamma, c \beta \gamma)
 \end{eqnarray}
 where 
 \[ x \equiv (ct,x) = (x_0,x) \]
 and Eqn(6.15) has been used to relate $d\tau$ and $dt$.
 Consider now a LT from S into another inertial frame S''
 moving with velocity $u$ along the x-axis relative to S. The
 corresponding spatial LT
 analagous to (3.3) is: 
 \begin{equation}
 x'' = \gamma_u(x-\beta_u x_0)
 \end{equation}
  where
 \[ \beta_u = \frac{u}{c}, ~~~\gamma_u = \sqrt{1-\beta_u^2} \] 
  Multiplying by $m$ and taking the derivative with respect to
  the proper time gives:
  \begin{equation}
 m\frac{dx''}{d \tau} = P''_x = \gamma_u(P_x-\beta_u P_0)
 \end{equation}
  It follows from Eqn(6.16) and (6.17) and the analagous
  four-vector definitions in S'' that:
  \begin{equation}
  P_0^2-P_x^2 = (P''_0)^2-(P''_x)^2 = m^2 c^2  
 \end{equation}
 Squaring Eqn(6.19), adding $ m^2 c^2$ to the LHS, using
 Eqn(6.20) and performing some algebra allows the
  derivation of the LT of the relativistic energy of
  O($m$):
  \begin{equation}
 P''_0 = \gamma_u(P_0-\beta_u P_x)
 \end{equation}
 Taking the ratio of Eqn(6.19) to Eqn(6.21):
  \begin{equation}
 \frac{P''_x}{P''_0} = \beta'' =\frac{\frac{P_x}{P_0}- \beta_u}
   {1-\beta_u \frac{P_x}{P_0}} = \frac{\beta-\beta_u}{1-\beta_u \beta}
 \end{equation}
 This equation may be rearranged as:
  \begin{equation}
 \beta = \frac{\beta''+\beta_u}{1+\beta'' \beta_u}
 \end{equation} 
 which is the usual relativistic velocity addition formula. It has been
 derived here on the basis of only the local LT between S and S' and
 the derivative with respect to the proper time of
  the  spatial LT between S and S''.   
Thus, the result is not dependent
 on the use of the LT of time Eqn(3.4), containing the term
 $-\beta \gamma x/c$ that is responsible, in the case of a non-local
 space-time  LT, for the unphysical `relativity of simultaneity'.
  Alternatively Eqn(6.21) can be more directly obtained by taking
 the derivative,
  with respect to the proper time, of the temporal LT between 
  S and S'':
  \begin{equation}
 x''_0 = \gamma_u(x_0-\beta_u x)
 \end{equation} 
  Because of the derivative, the choice of coordinate origins 
  is again of no importance. 
  \par In summary, the usual formulae of relativistic kinematics
  can be derived either by using only the spatial LT, 
  and a local LT to transform the
  time, or by taking derivatives with respect
  to proper time in both the spatial and temporal LT . No distinction
  between `local' and `non-local' LT is then required in the derivation 
  of kinematic formulae in SR. This conclusion may be more directly
  drawn from inspection of of the Minkowski plots in Fig.4. Relativistic
  kinematics concerns only timelike separated events and the corresponding
  hyperbolae are the same in Fig.4a (non-local LT) and Fig.4b (local LT).

\SECTION{\bf{The `Lorentz Fitzgerald Contraction' and the \newline
Michelson-Morley Experiment as a `Photon Clock'}}

  As discussed in most text-books on SR, the Lorentz Fitzgerald Contraction
  hypothesis was introduced by Fitzgerald~\cite{Lodge} and further developed
  by Lorentz~\cite{Lorentz} in an attempt to reconcile the null result
  of the Michelson-Morley (MM) experiment~\cite{MM} with the propagation
 of light as a wave motion in a luminiferous aether, through which
  the Michelson-Morley interferometer was conjectured to move with 
  velocity $v_{ae}$. The null result of the experiment is explained if,
  due to a dynamical interaction with the aether, the length of all
  moving bodies is contracted by the factor $\sqrt{1-(v_{ae}/c)^2}$.
  \par In Einstein's SR there is no aether, and the LC
   effect of the same size as that conjectured by Fitzgerald and Lorentz
   is found to be a geometrical consequence ($\Delta t = 0$ projection)
   of the space-time LT. The usual text-book discussion of the MM
  experiment, in terms of SR, claims to explain the null result by
  invoking a geometrical contraction of the arm of the interferometer
  parallel to the direction of motion by the Lorentz-Fitzgerald factor.
  Since this is often interpreted as experimental evidence for the LC,
  which it has been argued in the previous Sections of the present paper, 
   cannot be a real physical effect, it is mandatory to now re-examine
   carefully the description of the MM experiment in SR.
  \par Consider a Michelson interferometer with arms of equal
   length, $L$, at rest in the frame S'. The half-silvered plate is
   denoted by P and the mirrors by M1 and M2. The arm P-M2 is parallel
   to the direction of uniform motion of the interferometer, with velocity $v$,
    in the frame S (see Fig5a).
   At time $t' = 0$, a pulse of photons moving parallel to the x'-axis
   is split by P into two sub-pulses moving along the two arms of the
    interferometer. Typical photons moving in the arms P-M1 and P-M2 are
    denoted as $\gamma_1$ and $\gamma_2$ respectively. Each arm of the
   interferometer can now be considered as an independent `photon 
   clock' which measures the proper time interval in S' corresponding
   to the time for the photons to make the round trip from P to
   the mirrors and back again to P: $\Delta \tau = 2L/c$. First, the
   conventional argument~\cite{SK} for the existence of a LC effect for the 
   arm P-M2 will be examined, before analysing in detail the sequence 
   of events, as predicted by the LT, seen by an observer in S during
   the passage of the photons through the interferometer.
   \par The relativistic velocity addition formula (6.23) predicts that
   photons (or, in general, any massless particles) have the same velocity, $c$,
   in any inertial frame. Under the apparently plausible assumption (to be discussed
    further
    below) that the space-time events A (reflection of $\gamma_1$ from M1, 
     Fig.5b), B (reflection of $\gamma_2$ from M2, Fig.5c) and C (arrival of both
    reflected photons back at P, fig.5d) can be calculated by following 
    photon trajectories in the frame S, the following relations are derived:
 \begin{equation}
 c t_{OA} = \sqrt{L^2+v^2 t_{OA}^2} = c t_{AC} =
  \sqrt{L^2+v^2 t_{AC}^2} 
 \end{equation}
 \begin{equation}
 ct_{OB} = \ell + v t_{OB}, ~~~ ct_{BC} = \ell - v t_{BC} 
 \end{equation}
 Here $t_{OA}$ and $t_{AC}$ refer to time intervals along the 
  trajectory in S of $\gamma_1$, while $t_{OB}$ and $t_{BC}$ 
  are similarly defined for $\gamma_2$. The `observed length' of 
  the arm P-M2 is denoted by $\ell$. The total transit time
  in S of $\gamma_1$ is:
 \begin{equation}
 t_{OAC} =  t_{OA} + t_{AC} = \frac{2L}{c \sqrt{1-(\frac{v}{c})^2}} 
 \end{equation}
 while that of  $\gamma_2$ is      
 \begin{equation}
 t_{OBC} =  t_{OB} + t_{BC} = \frac{2\ell}{c (1-(\frac{v}{c})^2)} 
 \end{equation} 
 Assuming now that $\gamma_1$ and  $\gamma_2$ are observed in space-time coincidence 
   when they arrive back at C: $ t_{OAC} =  t_{OBC}$, it follows from Eqns(7.3)
  and (7.4) that:
 \begin{equation}
 \ell = L \sqrt{1-(\frac{v}{c})^2} 
 \end{equation}
 The `observed' length of the arm P-M2 is reduced,
  compared to its length in S', by the same factor as in
 the LC. However, the space-time observations performed here are quite
 different to the $\Delta t = 0$ projection that defines the LC. As will
  become clear when the pattern of space-time events seen by an observer
  in S is calculated in detail, at no point is a simultaneous
  observation made of both ends of the arm P-M2. Thus the SR `effect' 
  embodied in Eqn(7.5) is quite distinct from the LC, even though
  the length contraction factor is the same.
  \par The sequence of events observed in S during the passage of the 
   photons along the two arms of the interferometer is now considered. 
  It is imagined that each mirror, as well as the plate P, are equipped
   with small photon detectors, with a time resolution much smaller
   than $L/c$, that are used to trigger luminous signals in the
   close neighbourhoods of P, M1 and M2 so that the observer in S 
   observes space-time events corresponding to the passage of the
  photon pulses through P, their reflections from M1 and M2 and their
  return to P. It is also assumed that the observer in S records
  the actual (real) position of the interferometer for comparison with the apparent
   positions of the  
  signals generated by the passage of the photons.
\begin{figure}[htbp]
\begin{center}\hspace*{-0.5cm}\mbox{
\epsfysize10.0cm\epsffile{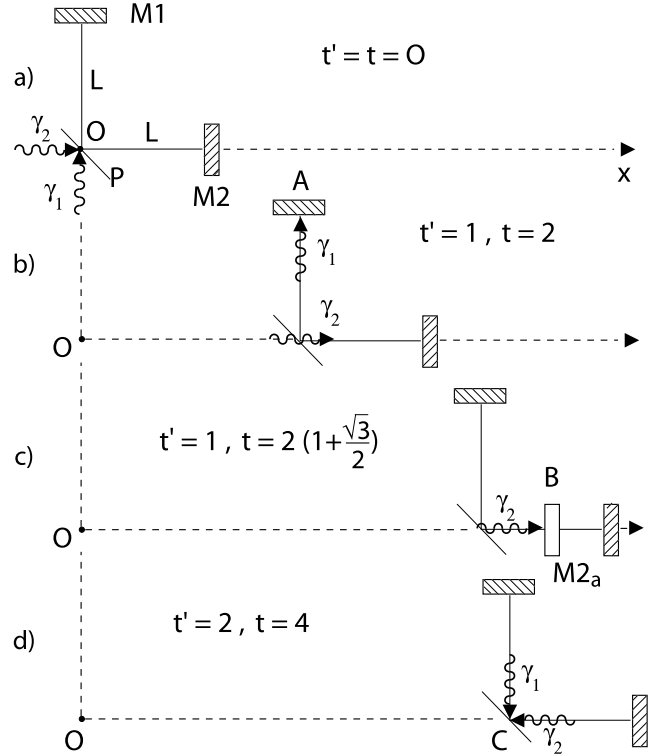}}
\caption{ {\em The sequence of events observed in S during the 
 passage of photon pulses through a Michelson interferometer moving 
 with constant velocity $\beta  = \sqrt{3}/2$. a) initial pulse arrives
 at the half-silvered plate, $P$, and divides into sub-pulses
  moving along the arms P-M1 ($\gamma_1$) and P-M2 ($\gamma_2$); b) $\gamma_1$
 reflects at M1 (event A);  c) $\gamma_2$  reflects at M2 (event B); 
 c) $\gamma_1$ and $\gamma_2$ return to $P$ (event C).
  The origin of S' is at $P$. Real positions of the moving mirrors are indicated
  by cross-hatched rectangles, apparent ones by open rectangles, or directly
  (M2$_a$).}}
\label{fig-fig5}
\end{center}
\end{figure}

\begin{figure}[htbp]
\begin{center}\hspace*{-0.5cm}\mbox{
\epsfysize10.0cm\epsffile{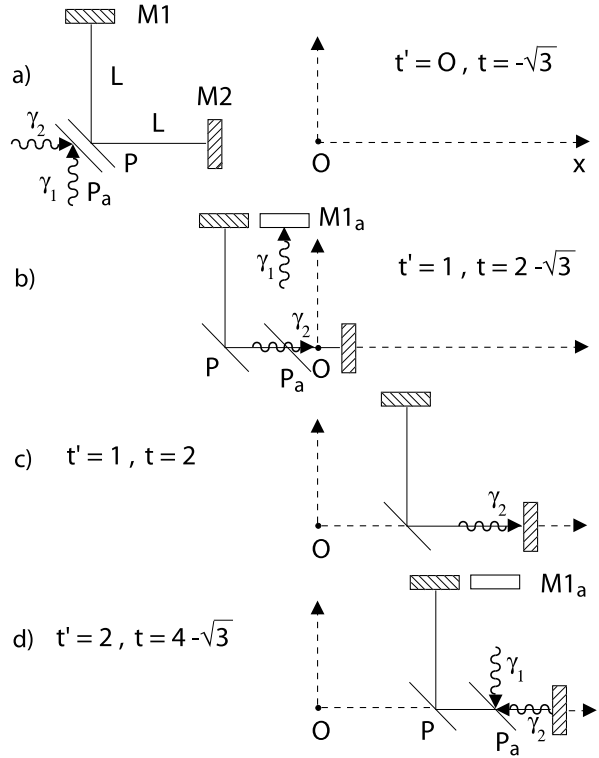}}
\caption{ {\em As Fig.5 except that the origin of S' is at M2}}
\label{fig-fig6}
\end{center}
\end{figure}

\begin{figure}[htbp]
\begin{center}\hspace*{-0.5cm}\mbox{
\epsfysize10.0cm\epsffile{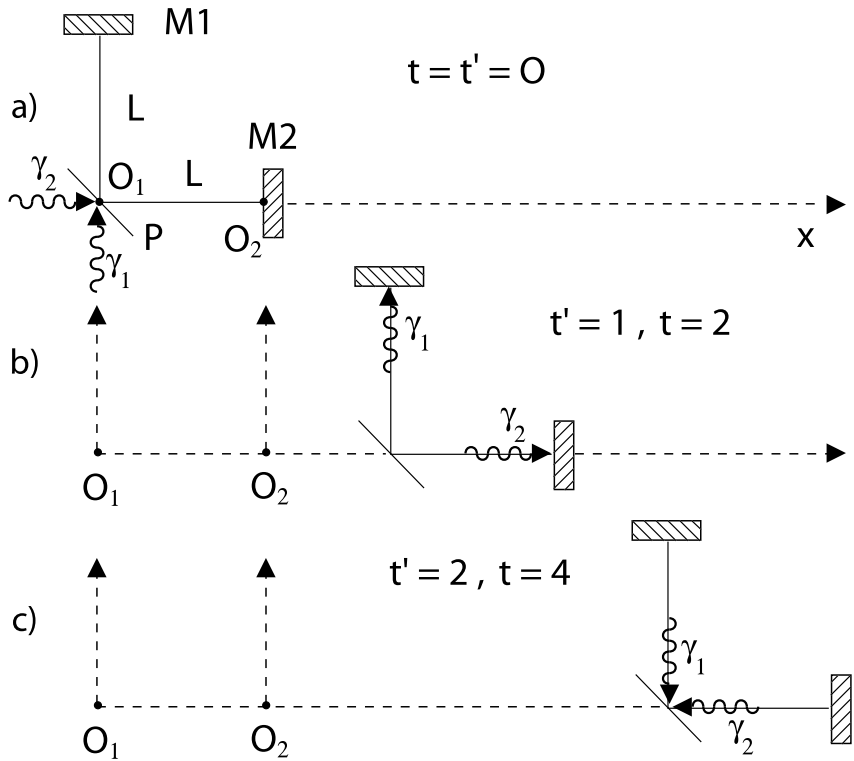}}
\caption{ {\em As Fig.5 except that local LT are used for all three
  events A, B and C}}
\label{fig-fig7}
\end{center}
\end{figure}

 \par The first case considered is a LT with origin in S' coincident
 with P. The LT is thus local for events situated  at P or M1 but
 non-local for events situated at M2. The positions and times in S'
  \footnote{In S' there is evidently no distinction between the real 
   and apparent positions of events},
 the apparent positions and times in S, as well as the corresponding
  real positions  of P of the four events: (i) initial passage of 
  photons through P, (ii) reflection of $\gamma_1$ from M1, (iii)
  reflection of $\gamma_2$ from M2 and (iv) return passage of
 $\gamma_1$ and $\gamma_2$ through P, are presented in Table 2 and shown 
  in Fig.5. In Figs.5, 6 and 7 the real positions in S of M1 and M2 are
  denoted by cross-hatched 
 rectangles and their apparent positions by open rectangles,
 \begin{table}
\begin{center}
\begin{tabular}{|c||c|c|c|c|c|c|} \hline
 Event   & $x'$ & $t'$  & $x^{app}$ & $t^{app}$ & $x_P^{app}$ &  $x_P$ \\
 \hline \hline
 $\gamma_1$, $\gamma_2$ pass P & 0 & 0 & 0 & 0 & 0 & 0 \\
       &  &  &  &  &  & \\
 $\gamma_1$ at M1 & 0 & $\frac{L}{c}$  & $\gamma \beta L$ &  $\frac{\gamma L}{c}$ &
  $\gamma \beta L$ &  $\gamma \beta L$ \\
       &  &  &  &  &  & \\
 $\gamma_2$ at M2 & $L$ & $\frac{L}{c}$  & $ \gamma(1+\beta)L$ & 
 $\frac{\gamma (1+\beta)L}{c}$ & $\gamma \beta (1+\beta) L$ &  $\gamma \beta L$ \\
       &  &  &  &  &  & \\
$\gamma_1$, $\gamma_2$ back at P & 0 & $2\frac{L}{c}$  &   $ 2 \gamma \beta L$   &
 $\frac{ 2 \gamma L}{c}$ &  $ 2 \gamma \beta L$  & $2 \gamma \beta L$ \\ 
       &  &  &  &  &  & \\     
\hline
\end{tabular}
\caption[]{{\em   Coordinates of space time events during
  photon transit of a Michelson interferometer. Origin in S' at P.
  The origin of S is at P at $t=0$}}      
\end{center}
\end{table}  
 \begin{table}
\begin{center}
\begin{tabular}{|c||c|c|c|c|c|c|} \hline
 Event   & $x'$ & $t'$  & $x^{app}$ & $t^{app}$ & $x_P^{app}$ &  $x_P$ \\
 \hline \hline
 $\gamma_1$, $\gamma_2$ pass P & $-L$  & 0 & $-\gamma L$ & $-\frac{\gamma \beta L}{c}$ &
  $-\gamma L$  &  $-(1+\gamma \beta^2)L$ \\
       &  &  &  &  &  & \\ 
 $\gamma_1$ at M1 &  $-L$  & $\frac{L}{c}$  & $-\gamma (1- \beta) L $ & 
  $\frac{\gamma (1-\beta) L}{c}$ &  $-\gamma (1- \beta) L $ & $-[1-\gamma \beta(1-\beta)]L$ \\
       &  &  &  &  & \\ 
 $\gamma_2$ at M2 & 0 & $\frac{L}{c}$  & $ \gamma \beta L$ & 
 $\frac{\gamma L}{c}$ & $-\gamma (1- \beta) L $ &  $-[1-\gamma \beta(1-\beta)]L$ \\
       &  &  &  &  &  & \\ 
$\gamma_1$, $\gamma_2$ back at P & $-L$  & $2\frac{L}{c}$  &  $-\gamma (1- 2 \beta) L $  &
  $\frac{\gamma (2-\beta)L}{c}$ &  $-\gamma (1- 2 \beta) L $  & $ ( 2 \gamma \beta-1-\gamma \beta^2)L $ \\ 
       &  &  &  &  &  & \\      
\hline
\end{tabular}
\caption[]{{\em Coordinates of space time events during
  photon transit of a Michelson interferometer. Origin in S' at M2.
  The origin of S is at distance $\gamma \beta^2 L$ from M2 at $t=-\gamma \beta L/c$.}}      
\end{center}
\end{table}
 \begin{table}
\begin{center}
\begin{tabular}{|c||c|c|c|c|c|} \hline
 Event   & $x'$ & $t'$  & $x^{app} = x_{source}$ & $t^{app}$ & $x_P$ \\
 \hline \hline
 $\gamma_1$, $\gamma_2$ pass P & 0 & 0 & 0 & 0 & 0 \\
       &  &  &  &  & \\ 
 $\gamma_1$ at M1 & 0 & $\frac{L}{c}$  & $\gamma \beta L$ &  $\frac{\gamma L}{c}$ &
  $\gamma \beta L$ \\
       &  &  &  &  & \\ 
 $\gamma_2$ at M2 & 0 & $\frac{L}{c}$  & $  L(1+\gamma \beta)$ & 
 $\frac{\gamma L}{c}$ &  $\gamma \beta L$  \\
       &  &  &  &  & \\ 
$\gamma_1$, $\gamma_2$ back at P & 0 & $2\frac{L}{c}$  & $ 2 \gamma \beta L$   &
 $\frac{ 2 \gamma L}{c}$ &  $ 2 \gamma \beta L$ \\ 
        &  &  &  &  & \\     
\hline
\end{tabular}
\caption[]{{\em Coordinates of space time events during
  photon transit of a Michelson interferometer. Local origins in S' at P (O$_1$')
 and M2 (O$_2$'). The origin, O$_1$, of S is at P at $t=0$. $x^{app}$ and $x_P$
   are relative to O$_1$. In this case there is no distinction, in the reference
   frame S, between the
    apparent positions of events and the real positions of their sources.}}      
\end{center}
\end{table} 
 and also directly
 (e.g. M2$_a$ in Fig.5c). The indicated positions of the plate P correspond to real 
  positions in S unless otherwise indicated (as P$_a$). The apparent positions in S
  of the photons at the indicated times are those of the tip of the
  corresponding arrowhead.
  The distances and times shown in Figs.5, 6 and 7 correspond, as previously, to the
  choice of parameters: $L = c =1$, $\beta = \sqrt{3}/2$. As shown in Fig.5a,b and d 
  the events (i), (ii) and (iv) are observed in spatial coincidence with the positions
   of P, M1 and, again, P, respectively. However, the events (ii) and (iii), that are
    simultaneous in S', are not so in S. The reflection of $\gamma_2$ on M2 is 
   observed at time $\gamma \beta L c$ later than the reflection of $\gamma_1$ from M1.
   Also the observed reflection from M2 is shifted from the actual position of M2 in S
   at the time of observation (Fig.5c). 
   \par The sequence of events presented in Table 3 and shown in Fig.6 results from
   choosing the origin of S' to be at M2. Now it is only event (iii) that 
   is observed in spatial coincidence with the actual position of the relevant 
   element of the interferometer (M2). Events (i), (ii) and (iv) are all observed at
   apparent positions separated from the actual positions of P, M1 and, again P, 
   respectively at the times of observation. In addition the whole sequence of
   events is shifted in time by $-\gamma \beta L/c$ as compared to those shown
   in Table 2 and Fig.5. This is the result of different clock synchronisation
    between S and S' in the two cases. For Table 2, the synchronised clock in S is
    situated at the position of P at $t=0$ (Fig.5a), whereas in Table 3 it 
    is at the position of M2 at $t=0$.
   \par The same critical remarks must be made concerning the different sequences
    of events predicted by the non-local LT in Figs.5 and 6, as previously made
    concerning the
    different apparent times and positions of the moving clocks in Figs.3b,c and d
    and Figs.4b,c,d and e, respectively.
    They correspond to physically impossible situations where different sequences
    of events are observed (with apparent positions of the events differently
     shifted from the actual positions of the mirrors and plate) depending only
   on an arbitary choice of the origin in S' of a non-local LT. As in the case of the 
    discussion of the LC in Section 4, the number of different possiblities is infinite.
    In the general case where the LT is non-local for all the events (i)-(iv), all of them
    will be observed at apparent positions different from the actual positions of P, M1 
    and M2 at the observation times in S. The sequences of events shown in Figs.5 and 6
    clearly violate the `CIPP' principle introduced in Section 4, as well as the
    `SSC' principle of Section 5. For example, in the case of Table 2 and Fig.5, 
      the photon reflection event and the mirror M2 are contiguous in
      space-time in S', but not, (as shown in Fig.5c) in S. Other examples of the 
       breakdown of SSC in the frame S are shown in Figs.6a, b and d.
    \par The results of using a local LT with origin in S' at P for events (i), (ii)
     and (iv), and at M2 for event (iii) are presented in Table 4 and shown in Fig.7. 
     In this case the CIPP principle does not apply since there is no arbitariness
     in the choice of coordinate origins, and in all cases SSC is respected. 
     Transmission events and the position of P, and reflection events and the
     positions of the mirrors, are contiguous in space-time in both S' and S.
     In this case, there is no apparent contraction of the arm P-M2 as the
     local LT leaves invariant spatial separations. 
     It is conjectured here that if the gedankenexperiment just discussed were to be
     actually performed, the observations would be as in Table 4 and Fig.7. 
     Indeed, in the case of a non-local LT there is no definite prediction to
     be tested, since what are shown, in Tables 2 and 3, and Figs.5 and 6, are
      just two of an infinite
     number of predictions in each of which apparent events are observed
     at different positions, relative to the actual positions of the 
     interferometer in S.   
     \par It should be remarked that the simple calculation, based on the 
       hypothesis of an apparent velocity in S of $c$, recapitulated above,
       that is used to infer length contraction of the arm P-M2, is consistent
       with the sequence of events shown if Figs.5 and 6, and is indeed so for
       any choice of origin in S' of the non-local LT. The apparent 
      contraction of the length of the arm P-M2 by the factor $1/\gamma$ is
     evident in Fig.5c and Fig.6b and d. The situation is similar to that 
     shown in Fig.3b,c,d and e where, although the non-local LT predicts
     different apparent positions of the clocks, the apparent
     distance between the clocks is constant and shows the LC effect. 
   As already mentioned, it is clear from the sequences of events
    presented in Tables 2 and 3 that although the size of the apparent
    contraction of the arm P-M2 is the same as for LC (the $\Delta t = 0$ 
     projection of the LT) its origin is completely different, since none of
    the events listed correspond to simultaneous observation of P and M2 from S.
   \par It will now be found instructive to consider in more detail the
        two independent `photon clocks' constituted by the two arms of the
       moving interferometer. That corresponding to the arm P-M1 may be called
    a `transverse' $\gamma$-clock, that to P-M2 as a `longitudinal' one. By 
     comparing the events listed in Tables 2, 3 and 4, it can be seen that   
     the transverse $\gamma$-clock gives a similar temporal sequence of events for
     a non-local or local LT. The transit times from O to A and from A to B (see Fig.5)
      are equal and given by Eqn(7.1). Direct calculation using the coordinates 
      presented in Tables 2, 3 and 4 shows that the apparent distances in S
      between events (i) and (ii) and between (iii) and (iv), are, in all cases,
       $\gamma L$, while the corresponding elapsed time is $\gamma L/c$, thus 
       yielding an apparent velocity of $c$, consistent with the assumption
     used to derive Eqn(7.1). The behaviour of the transverse $\gamma$-clock therefore
     does not depend on the choice of non-local or local LT to analyse its 
     behaviour. The situation is different for the longitudinal $\gamma$-clock
     based on the arm P-M2. As can be seen from Tables 2 and 3 (the same is true for
     any other origin in S' of the non-local LT) and already mentioned above,
     the event (iii) is observed to occur in S at time $\gamma \beta L/c$ later
     than (ii). Also the apparent velocity in S of the photon $\gamma_2$ in both
       the path
      OB and BC is always $c$, and so is consistent with Eqn(7.2) above. When 
     a local LT is used to transform every event from S' to S, as in Table 4,
     different apparent velocities are found for the paths OB and BC:
     \begin{eqnarray}
      v^{app}_{OB} & = & c\left(\frac{1}{\gamma}+\beta \right) \\
      v^{app}_{BC} & = & c\left(\frac{1}{\gamma}-\beta \right)
     \end{eqnarray}
      Since the total apparent path length according to Table 4 is $2L$
    and the total elapsed time in S is $2 \gamma L/c$ the average apparent 
    velocity in the arm P-M2, $\langle v^{app}_{long}\rangle$ is:
     \begin{equation}
 \langle v^{app}_{long} \rangle = \frac{c}{\gamma}
     \end{equation}
     Since a local LT modifies time intervals but not spatial separations
      when transforming events between two inertial frames, and as the 
      apparent velocity is just the ratio of the spatial separation
     to the time interval, Eqn(7.8) is  natural consequence of the TD effect.
      It can be seen from Eqns(7.6), (7.7) and (7.8)
     that $ v^{app}_{OB} > c$, $ v^{app}_{BC} <  c$ and $\langle v^{app}_{long}\rangle < c$.
     It may seem, at first sight that Eqns(7.6) and (7.7) are in contradiction with 
     the relativistic velocity addition formula (6.23). However the meanings 
     of these equations are completely different. Eqn(6.23) describes the relation
      between velocities of some physical object as measured in three different
      inertial frames,
         whereas
      Eqn(7.6) and (7.7) describe the observation, from another
      inertial frame, of a clock that uses 
       the constancy of the velocity of light propagating over a 
      known fixed distance
      to define a time interval.
    \par Finally, in this Section, it may be remarked that Eqn(7.8) has a 
     simple physical interpretation in the case that all space-time 
     events are transformed locally between S' and S. When a local LT is used for a moving
      macroscopic
     clock (of any type) the movement of all space points of the
     clock is apparently slowed down by the universal factor $1/\gamma$. 
     Compare the longitudinal $\gamma$-clock of this type where the photon
     makes a round trip from P to M2 and back to P, with a conventional
     analogue clock, where the hands make a single rotation, thus also
      returning to the original configuration. Just as the rate of    
   rotation of the hand of the clock is apparently slowed down by the
      fraction $1/\gamma$, due to TD, so too is the apparent speed of the
      photon in the $\gamma$-clock slowed down by the same factor
      over one complete period of operation.

\SECTION{\bf{A Summary of Experimental Tests of Special Relativity}}

  In this Section a brief review is made of the experimental tests
  of SR that have been performed to date. They can be grouped under
  four broad categories: (i) tests of Einstein's second postulate
  (universality, isotropy and source motion independence of the
   velocity of light) (ii) measurements of the Transverse Doppler
   Effect, (iii) tests of relativistic kinematics and (iv) tests of   
   Time Dilatation. Some more general (but model dependent) tests
   based on electron and muon $g-2$ experiments are also briefly
   mentioned. Finally, possible indirect evidence for LC derived
    from models of particle production in
   nucleon-nucleon collisions is discussed.
   \par An attempt has been
   made to be exhaustive concerning the different types of test 
   that have been performed, but not as regards the detailed
   literature for each type. As far as possible, the most recent
   and precise limits are quoted.

   \par {\bf \Large Isotropy and Source Independence of the Velocity
          of Light}
    \par The first experiment of this type, already discussed in the 
     previous Section, is that of MM~\cite{MM}. The limit obtained for the 
     maximum possible velocity, $v_{ae}$, of the Earth relative to the aether
     of about $5$ km/sec is much less than either the Earth's
     orbital speed around the Sun (30 km/s) or the Earth's velocity 
     relative to the rest frame of the microwave background radiation
     (380 km/s). However, because the MM interferometer had arms of
     equal length, the `aether drag' Lorentz Fitzgerald Contraction
     hypothesis could not be excluded. This was done by the
     Kennedy-Thorndike (KT)~\cite{KT} experiment
     which was a repetition of the MM experiment with
     unequal length arms. In more recent years the use of laser
     beams with frequencies servo-locked to standing-wave modes
     of a Fabry-Perot interferometer of precisely controlled
     length~\cite{BH,HH} has allowed a 300 fold improvement
     in the limit on a possible anisotropy of the velocity of
     light from $\Delta c/c \leq 3 \times 10^{-8}$ in the original
     KT experiment to  $\Delta c/c \leq 10^{-10}$~\cite{HH}.
     \par One of the earliest suggestions to test the independence of
      the velocity of light from the motion of its source was that of
      De Sitter~\cite{deSitter} who suggested observation of the 
      periodic variation of the Doppler effect in the light from
      binary star systems. If the velocity of light depends on 
      that of the source according to the classical velocity addition
      formula, the half-periods observed using the Doppler effect
      of each member of the binary system are predicted to be 
      unequal~\cite{Aharoni}. It was pointed out by Fox~\cite{Fox}
      that if the light is scattered from interstellar matter before
      arriving at the Earth, then 
      due to the optical `Extinction Theorem', the source of the
      observed light is, effectively, the scattering atoms, not the moving stars,
      thus invalidating the test. This objection is not applicable to
      photons in the X-ray region due to their much smaller scattering
      cross-section. An analysis~\cite{Brecher} of pulsating X-ray sources set a limit of
      $k < 2 \times  10^{-9}$ in the modified classical velocity addition 
      formula: $c' = c + kv$. An accelerator experiment, in
      which the velocity of photons originating in the decay of $\pi^0$ mesons
      of energy about 6 GeV was measured by time-of-flight, set the limit 
      $k = (-3 \pm 13) \times  10^{-5}$~\cite{AFKWB}. The latter two 
      experiments are, then, essentially tests of the relativistic velocity
      addition formula (6.23).

    \par {\bf \Large Measurements of the Transverse Doppler Effect}
      
      \par Following the pioneering experiment of Ives and Stilwell~\cite{IS}
        using a beam of atomic hydrogen, the Transverse Doppler Effect
        (essentially the prediction corresponding to the first term on the
         RHS of Eqn(6.21)) has been measured using many different experimental
       techniques: temperature dependence of the M\"{o}ssbauer effect~\cite{PR},
      variation of the absorbtion of photons from a rotating M\"{o}ssbauer
      source~\cite{HSCE,Kundig}, double photon excitation, by a single laser,
      of beams of neon~\cite{KPEL,MGSS} or
      collinear saturation spectoscopy using a  $^7\rm{Li}^+$ beam with 
      $\beta = 0.064$ and two independent laser beams ~\cite{Griesereta} and
      laser excitation of an 800 MeV ($\beta = 0.84$) atomic hydrogen 
      beam~\cite{MacArthureta}. The two photon excitation experiments
      measure the pure Transverse Doppler Effect. In Reference~\cite{KPEL}
       this term was measured with an accuracy of $4 \times 10^{-5}$ for
      $\beta = 4 \times 10^{-3}$. In a later experiment of the same
      type~\cite{MGSS}, the precision was improved to $2.3 \times 10^{-6}$.
       The experiment of Reference~\cite{Griesereta} reached a similar precision
       of $1.1 \times 10^{-6}$.
      The experiment of Reference~\cite{MacArthureta} is interesting both
      because of the highly relativistic nature of the atomic Hydrogen 
      beam ($\gamma = 1.84$) in comparison with the atomic beams used in
      References~\cite{KPEL,MGSS,Griesereta} and that, not only
      the Transverse Doppler Effect, but, in fact, the whole relativistic
      energy transformation equation Eqn(6.21) is tested with a relative 
      precision of $2.7 \times 10^{-4}$. 

   \par {\bf \Large Tests of Relativistic Kinematics}
   
     As just discussed, relativistic kinematics is already tested, using photons,
     in the Transverse Doppler Effect experiments. The relativistic kinematics
     of the electron embodied in the relation $p = \gamma m v$ provided  provided 
     one of the first pieces of evidence~\cite{Bucherer} for the correctness of
     the kinematical part of Einstein's theory of SR.
     More recent tests of relativistic kinematics
     using electrons have included measurements of the relation between kinetic
     energy and velocity~\cite{Bertozzi} or kinetic energy and momentum
     ~\cite{Parker,GK}. The first of these experiments used a Van de Graff accelerator
     to produce an electron beam with kinetic energies from 0.5-15 MeV, whose velocity
     was subseqently measured by time-of-flight. The latter two experiments used
     radioactive $\beta$-decay sources, a magnetic spectrometer to measure the
     momentum and total absorption
      detectors to measure the kinetic energy. An experiment performed at SLAC measured,
      by time-of-flight, the velocity of 15-20 GeV photons and electrons over a distance
      of $\simeq 1$km. The fractional velocity difference was found to be zero within
      $2 \times 10^{-7}$~\cite{GRYGM}. This is essentially a test of the relation
      between energy momentum and mass (Eqn(6.20) as well as that between velocity,
      momentum and energy: $\beta = pc/E$. 
 
  \par {\bf \Large Tests of Time Dilatation}

    \par  Precise test of TD have been performed either by using decaying subatomic particles
    as `radioactive clocks', as described in Section 3 above, or by comparing an atomic clock,
     at rest on the surface of the Earth, with a similar clock in movement (or that
     has been in movement) in an aircraft, a ballistic rocket or a satellite moving
     around the earth.
    \par In the experiment of Ayres et al.~\cite{Ayresetal} the measured lifetimes
      of 300 MeV ($\gamma = 2.45$) $\pi^{\pm}$ produced at a cyclotron were compared
      with the known lifetime of pions at rest. The velocity of the pions, measured
      by time-of-flight, was used to calculate $\gamma$. The time dilated lifetime
      was found to agree
      with the prediction of SR to within  $4 \times 10^{-3}$. 
    \par Bailey et al.~\cite{Baileyetal} measured the lifetimes of 3.1 GeV ($\gamma = 29.3$)
     $\mu^{\pm}$ decaying in the storage ring of the last CERN g-2 experiment. The ratio of 
     the muon lifetimes in flight $\tau^{\pm}$ to the known values at rest $\tau_0^{\pm}$
     were compared with the quantity $\overline{\gamma}$ derived from the measured 
      cyclotron frequency of the stored muons. For $\mu^+$, where the lifetime at rest
     is known with the best accuracy, the ratio $\tau^+/\tau_0$ was found to agree
     with $\overline{\gamma}$ to within  $(2 \pm 9)  \times 10^{-4}$. The physical 
     meaning of this test is discussed in more detail below. Another important feature 
     of this experiment is the demonstration that the TD effect depends only on
     the absolute value of the velocity. The measured effect is the same as in an
     inertial frame even though the transverse acceleration of the circulating 
     muons is  $ 10^{19}$ m s$^{-2}$.

    \par In the experiment of Hafele and Keating~\cite{HK} four cesium clocks were 
     flown around the world in commercial airliners, once from west to east (WE) and
     once from east to west (EW). They were then compared with similar reference
     clocks at the U.S. Naval Observatory, thus realising the `twins paradox' of
     SR in an actual experiment. The effects of Special Relativity (TD) and of
     General Relativity (the gravitational red-shift) were predicted to be of
     comparable size, but to add on the EW and subtract on the WE trips. The 
     observed time losses of the flown clocks relative to the reference clocks:
     $59 \pm 10$ ns (WE) and  $273 \pm 7$ ns (EW) were found to be in good 
     agreement with the relativistic predictions of $40 \pm 23$ ns and
      $275 \pm 21$ ns respectively.      
     \par A combined test of SR and General Relativity (GR) was also made in the
      experiment of Vessot et al.~\cite{Vessotetal} in which a hydrogen maser 
      was flown in a rocket executing a ballistic trajectory with a maximum
      altitude of $10^4$ km. Again the SR and GR corrections affecting the observed
      rate of the maser were of comparable magnitude and opposite sign. They were seen
     to cancel exactly at a certain time during both the ascent and descent of the
      rocket. The quoted accuracy of the combined SR and GR test was quoted as
     $(2.5 \pm 70)  \times 10^{-6}$
       \par In the Spacelab experiment NAVEX~\cite{Sappl}, a cesium clock in a Space
     Shuttle was compared with a similar, ground based, clock. The difference
     in rate $R$ of the two clocks is predicted to be:
      \begin{equation}
      R = \frac{1}{c^2}(\Delta \phi - \frac{\Delta v^2}{2})
     \end{equation}
   where $\Delta \phi$ is the difference in gravitational potential and 
    $\Delta v^2$ the difference in the squared velocities between the moving 
     and reference clocks. The measured value of R: $-295.02 \pm 0.29$ ps/s
    was found to be in good agreement with the relativistic prediction:
    $-295.06 \pm 0.10$ ps/s. The GR part of the prediction for R:
    $R_{GR} = 35.0 \pm 0.06$ ps/s is thus tested with a relative precision of about
    1$\%$, and the SR part (TD) with a relative precision of 0.1$\%$.         
    In this experiment the Shuttle was orbiting at $\simeq 330$km above the Earth
    at a velocity of 7712 km/s ($\beta =  2.5  \times 10^{-5}$). As discussed in
     the following Section, a straightforward extension of this type of 
     experiment could test the relativity of simultaneity of SR.
     Such a test has never been performed.
 
  \par {\bf \Large Relativity Tests in Lepton g-2 Experiments}
     
   \par The rate of precession, $\omega_a$, of the spin vector of a charged
    lepton, relative to its momentum vector, in a uniform magnetic field, is
    independent of the energy of the lepton. Closer examination shows that
    this fact results from a subtle cancellation of several SR effects in the 
    three angular frequencies that make up $\omega_a$: the Larmor precession
     frequency, observed in the laboratory frame, $\omega_L$,  the Thomas
      precession frequency, $\omega_T$, and the cyclotron frequency, $\omega_c$,
     In fact~\cite{CFFP}:
     \begin{eqnarray}
   \omega_a & = & \omega_L+\omega_T-\omega_c \nonumber \\
            & = & \frac{eB}{mc}  \left[\frac{g \gamma_E}{2 \gamma_t}
    +\frac{1-\gamma_T}{\gamma_M}
           -\frac{1}{\gamma_M}\right] 
   \end{eqnarray}
  where the 3 terms in the large square bracket correspond to  $\omega_L$,
    $\omega_T$ and  $\omega_c$ respectively. The four different $\gamma$-factors
    $\gamma_E$, $\gamma_t$, $\gamma_T$ and  $\gamma_M$ correspond to the
    Lorentz transformations of the electromagnetic field and time, to the
    kinematical Thomas precession: ( $(1-\gamma_T)\omega_c$ ) and the relativistic
    mass-energy relation: ( $\gamma_M = E/(mc^2)$ ), respectively. The quantity
     $\overline{\gamma}$ refered to above, as determined from the cyclotron 
      frequency of the CERN g-2 experiment~\cite{Baileyetal} is, in fact,
    $\gamma_M$, while the ratio of the muon lifetimes at rest and in flight
     is $\gamma_t$. Thus, in this experiment, the relation $\gamma_t =\gamma_M$
     is checked. In the pion decay experiment of Ayres et al.~\cite{Ayresetal}
     the relation tested is  $\gamma_t =\gamma_v$ where $\gamma_v \equiv 
     \sqrt{1-(v/c)^2}$, since the velocity of the deaying pions is directly
     measured. To make further tests, based directly on the measurement
    of $\omega_a$, additional assumptions are necessary. For example,
      Newman et al.~\cite{NFRS} assume that $\gamma_t = \gamma_E
      = \gamma_T \equiv \gamma_k$, since they are all `kinematical' 
      quantities, but that $\gamma_M$ might be different. By comparing
      measurements of  $\omega_a$ for the electron at velocities
     $\beta \simeq 0.5$ and  $ 5  \times 10^{-5}$ it is concluded.~\cite{NFRS}
     that $ \gamma_k$ and $\gamma_M$ are equal to within a fractional error
     of  $ (5.3\pm 3.5)  \times 10^{-9}$, yielding the most precise experimental
     confirmation of SR. Combley et al.~\cite{CFFP} test SR using Eqn.(8.2)
     but with considerably weaker theoretical assumptions than Newman et al. By 
      considering the transverse acceleration of a charged particle 
     in a magnetic field in both the laboratory and particle rest frames,
    the consistency condition:
     \begin{equation}
       \gamma_M  \gamma_E =  \gamma_t^2
     \end{equation}
     may be simply derived. Combining Eqns(8.2) and (8.3) with the direct
     measurements of $\gamma_t$ and $\gamma_M$ from the CERN muon g-2
     experiment both $\gamma_E$ and $\gamma_T$ were derived. All four
       $\gamma$-factors are found to be consistent, within their 
      experimental errors, the largest of which is
      $ 2  \times 10^{-3}$~\cite{CFFP}.

  \par {\bf \Large Test of Length Contraction with Particle Production Models}

  \par In the statistical model of particle production in high
  energy nucleon-nucleon collisions proposed by Fermi~\cite{Fermi1,Fermi2}
  and the related hydrodynamical model of Belenkij and Landau~\cite{BL},
  the initial state is supposed to consist of a length-contracted
   `pancake' of high density nuclear matter (see, for example, Fig.1
   of Reference~\cite{Fermi2}) which then develops into the observed 
   multiparticle final state. The multiplicity of produced 
   particles is then expected to be different in such models
   according to whether the volume, V, of the initial state takes into
  account, or not, LC. Fermi remarked~\cite{Fermi2} that the agreement
   of the
  model with experimental measurements `seems to indicate that 
  the assumption that the volume, V, should be Lorentz contracted
  is not greatly in error.' However, the following note of caution
  was added:` One should keep in mind, however, that the number
  of particles emitted in a collision of this type depends only
  on the fourth root of the volume V. A change of a factor of 2 or 3 
  would produce only a relatively minor variation in the expected
  number of particles'. In a later review article
  Feinberg~\cite{Feinberg} re-examined the question as to whether
  the comparison of such models with experiment could be interpreted
  as providing positive evidence for LC, and concluded that this 
  was not the case. With our present knowledge of the quark 
  substructure of nucleons and of QCD the physical basis of
  the models of Fermi and Landau may, of course, be questioned.
  The present writer concludes, finally, on this question, that statistical
  and hydrodynamical models of particle production do not provide
  any convincing experimental evidence for the existence, or not, of the 
  LC as a physical effect.

 \SECTION{\bf{A Proposal for an Experiment in Space to Test the
            Relativity of Simultaneity}}

 The review of experimental tests of SR presented in the previous Section
 shows that, at the time of writing, although there is ample experimental
 confirmation of Einstein's second postulate, relativistic kinematics
 and time dilatation, this is not the case for length contraction and
 the other apparent space-time effects: time contraction and
   space dilatation~\cite{JHF1}.    
 This disparity in the experimental verification of SR is not at
 all reflected in text book discussions where no discrimination
  is made between the well tested TD effect, and the experimentally untested LC
  effect. To the writer's best knowledge, no experiment has 
  even attempted to observe LC or the relativlty of simultaneity,
  much less the recently noted TC and SD effects that are also
  direct consequences of the relativity of simultaneity. In this Section
  a test of the relativity of simultaneity using similar techniques
  to the previously performed Spacelab experiment NAVEX (SEN)~\cite{Sappl},
  is proposed.
  It is essentially a method to observe the previously proposed~\cite{JHF1}
  TC effect.
  \par The principle of the experiment is illustrated by considering observation
 of the two clocks A and B, introduced in Section 2 above, from the
  fixed position, in the stationary frame S, of the clock C. The clocks A and B
  are separated by the fixed distance $L$ in their common rest-frame, S', and
  are synchronised (see Figs.1a, 8a and 9a). The frame S' moves with uniform
  velocity $v$ along the x-axis. Clock C is synchronised with B at time $t =t'=0$
  when B has the same x coordinate in S as C. The results of the conventional
  SR calculation\footnote{This is simliar to Einstein's calculation of
  LC in Ref~\cite{Einstein2}}, choosing the origin in S' at the position
  of clock B, and so using a non-local LT for the clock A, are shown in Table 5
  and Fig.8. The results of using a local LT for both clocks A and B are shown in 
  Table 6 and Fig.9. In the case of the non-local LT for A the apparent
  position of this clock coincides with that of C at time
  $t_c = T_{NL} = L/\gamma v$ when the time , $t'^{app}_A$, indicated by A
  when viewed from S, is given by $t'^{app}_A = L/v$. Thus $t'^{app}_A = \gamma T_{NL}$.
   The moving clock A thus appears to be {\it in advance} of the
  stationary clock C by $\gamma-1$ times the time interval in S, $T_{NC}$, between 
  the passages of the clocks B and A past C. This is just the TC effect
  ($\Delta x = 0$ projection of the LT) pointed out in Reference~\cite{JHF1}.
   Because clock C is synchronised with B at $t=0$ it is easy to see that, unlike 
   the apparent positions of the clocks discussed in Section 4, the TC effect 
   is the same for any choice of the origin of the non-local LT in S'. This 
   follows (see Figs.2b,c,d and Table 1) because the relativity of simultaneity
   gives always gives the same apparent time difference 
    $t'_A-t'_B = \beta L/c$ between the times indicated by A and B in S. Thus,
    unlike in the case of the prediction of the apparent positions of the
    clocks, the TC effect is unambigously predicted for any choice of the origin
    of the non-local LT, and so is experimentally testable.

\begin{figure}[htbp]
\begin{center}\hspace*{-0.5cm}\mbox{
\epsfysize10.0cm\epsffile{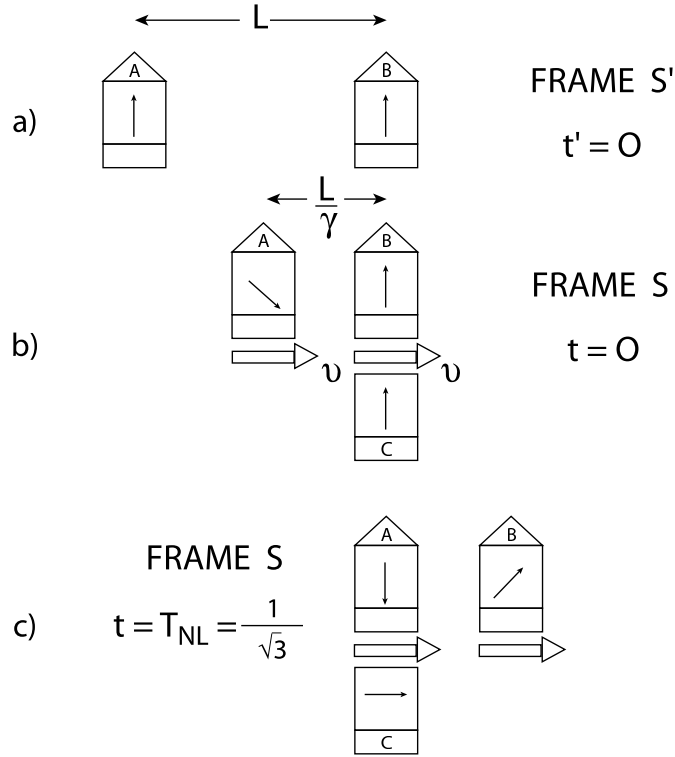}}
\caption{{\em Clocks A and B, at rest in S', move with velocity $v = \frac{\sqrt{3}}{2}c$ along
  the x-axis in S. When B passes the fixed clock in S, these clocks are
  synchronised ($t = t' = 0$), as are A and B in S'. The times of A and B
  as observed in S' at $t' = 0$ and in S at  $t = 0$ are shown in a) and
  b) respectively. c) shows the times indicated by the clocks when the
  apparent  x-position of A coincides with the x-position of C. The origin
  of S' is at B, i.e. the LT is non-local for clock A.}}
\label{fig-fig8}
\end{center}
\end{figure}

 \begin{table}
\begin{center}
\begin{tabular}{|c||c|c|c|c|c|c|} \hline
 Event   & $x_C$ & $t_C$  & $x^{app}_A$ & $t'^{app}_A$ & $x^{app}_B$ & $t'^{app}_B$ \\
 \hline \hline
 & & & & & & \\
B passes C &  0 & 0 &  $-\frac{L}{\gamma}$ &  $\frac{ \beta L}{c}$ & 0 & 0 \\
 & & & & & & \\
A passes C &  0 &  $\frac{L}{\gamma \beta  c}$ & 0 &  $\frac{L}{ \beta c}$ &
  $\frac{L}{\gamma}$  &  $\frac{L}{ \gamma^2 \beta c}$ \\ 
 & & & & & & \\  
\hline
\end{tabular}
\caption[]{{\em Coordinates of space time events in S and S'. The origin
   of S' is at clock B, so that the LT is non-local for clock A. The origin
  of S is at C at $t_C = 0$.}}      
\end{center}
\end{table}
  
\begin{figure}[htbp]
\begin{center}\hspace*{-0.5cm}\mbox{
\epsfysize10.0cm\epsffile{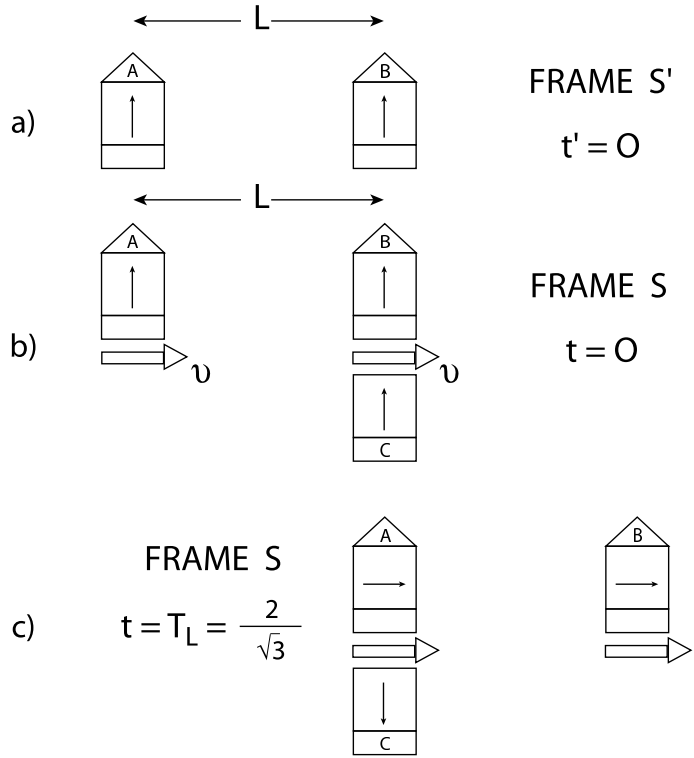}}
\caption{{\em As Fig. 8, except that local LT are used for both A and B.}}
\label{fig-fig9}
\end{center}
\end{figure}

   \par Use of local LT for both A and B gives the results (respecting translational
    invariance): $t'_A = t'_B$ (for any $t_C$) and  $t'^{app}_A = T_{L}/\gamma$.
    Thus clock A appears to be {\it delayed} relative to the stationary clock,
    by $(\gamma-1)/\gamma$ times the time interval in S, $T_L$, between 
    the passages of the clocks B and A past C. Also, since there is no LC effect
    in this case, $T_L = \gamma T_{NL} = L/v$.
     \par In SEN a cesium clock (that may be
      identified with the clock B above) in a Space Shuttle, executing an almost
      circular orbit around the Earth, was compared with a similar ground-based
      clock (corresponding to clock C above) at the culmination times of the 
      orbit, i.e. the times at which the distance between the orbiting and 
      ground based clocks was minimum. In order to realise the experiment
      described above, it is sufficient to add a third clock 
     (corresponding to A in the above example) following the same orbit as B
     but separated from it by a distance $L$. Comparison of A and C at 
     culmination, having previously synchronised B and C at culmination,
     then essentially realises the ideal experiment discussed above and so 
     enables a test of the relativity of simultaneity of SR.
 
 \begin{table}
\begin{center}
\begin{tabular}{|c||c|c|c|c|c|c|} \hline
 Event   & $x_C$ & $t_C$  & $x_A$ & $t'^{app}_A$ & $x_B$ & $t'^{app}_B$ \\
 \hline \hline
 & & & & & & \\
B passes C &  0 & 0 &  $-L$ &  0 & 0 & 0 \\
 & & & & & & \\
A passes C &  0 &  $\frac{L}{\beta c}$ & 0 &  $\frac{L}{\gamma \beta c}$ &
  $L$  &  $\frac{L}{\gamma \beta c}$ \\ 
 & & & & & & \\  
\hline
\end{tabular}
\caption[]{{\em Coordinates of space time events in S and S'. Local LT
  are used for both clocks A and B. The origin
  of S is at C at $t_C = 0$.}}      
\end{center}
\end{table}  

      \par To estimate the order of magnitude of the expected effects and the 
      corresponding measurement uncertainties, the movement of the ground
     based clock due to the rotation of the earth is, at first, neglected.
     It is also assumed that the clocks A and B are in the same inertial
     frame. For this discussion, the orbit parameters of SEN~\cite{Sappl}
     will be assumed. The Space Shuttle was in circular orbit 328 km
     above the surface of the Earth, moving with a velocity of
      7712 km/s ($\beta = 2.5 \times 10^{-5}$).
     A convenient separation of A and B along their common orbit is then
     200km. In this case synchronisation signals can be exchanged 
     between these clocks above the Earth's atmosphere. Thus $T \simeq
       T_L \simeq T_{NL} = 26$ s. Since, to first order in $\beta^2$,
      $\gamma -1 = (\gamma-1)/\gamma = \beta^2/2$, the usual SR calculation,
       using a non-local LT for the clock A, predicts that it will be
      observed to be {\it in advance} of C by $26 \times  \beta^2/2 = 8.1$ ns,
     whereas the calculation with a local LT for both A and B predicts 
      a {\it delay} of the same size. The latter is just the universal TD effect
      for all clocks at rest in S', that is required by translational
      invariance. The time shifts relative to C are proportional to
      $T$, and so to the separation of A and B along their common orbit.
      The time shift should be easily measurable in even a single 
      passage through culmination of clocks B and A. For example, in SEN, the
      quoted experimental precision on the combined (SR+GR) relativistic
      corrections to the rate of the moving clock corresponded to an 
      experimental time resolution of about 0.1 ns in the time difference
      of clocks B and C over the rotation period of the Shuttle (1.6 h). 
      The intrinsic uncertainty in the relative rates of B and C over a 
      period of 26 s, using clocks of the same type as used in SEN, would
      contribute an uncertainty of only $\simeq 7.5$ ps.
      \par The gravitational red shift, due to the Earth's field, has the 
       effect of increasing the advance, (or decreasing the delay) of A,
       relative to C, by about 10$\%$ for the case of a non-local, (or local) 
       LT applied to A.
      \par The LC effect for the non-local LT of A results in a smaller 
       value of $T$: $T_{NL} = T_L/{\gamma} = T_L(1-\beta^2/2 +O(\beta^4))$.
       In principle, this difference is measurable if the absolute distance, $L$,
       between A and B in S' is precisely known, as well as the time 
       difference between the culminations of A and B and the velocity
       of the orbiting clocks. Although the first 
       condition can perhaps be met by using interferometry, (the LC
        effect corresponds to a difference of length of $\simeq 60~\mu m$ 
       over 200 km) the rotation of the Earth would seem to preclude any
       possiblity to measure the latter quantities with the required 
       uncertainty in time of about 1 ns in 30 s, and knowledge
       of the orbit velocity with a similar precision. The reason for this is
       that the
       movement of C on the surface of the Earth between the culminations of
       A and B, is expected to modify $T$ by about 4 $\%$; while A is moving
       the distance of 200 km so as to occupy the position of B at culmination,
       the clock C moves about 8.5 km due to the Earth's rotation. This is not
       serious for the tests of TC and TD since the expected time shifts are,
       in first approximation, simply scaled acording to the actual value of
       $T$. However,the spatial separation between the culminations of A and
       B is clearly quite different to the ideal case (i.e., neglecting the
       rotation of the Earth). The separation must be known with a 
       relative precision of $\simeq 3 \times 10^{-10}$ in order to test
       directly the LC effect. This hardly seems possible.

 \par {\bf Added Note}
 \par  In Section 3, Eq.~(3.7) is incorrect and should be replaced by: $dt' = dt/\gamma(v)-(x'/c)d\beta$. In
  consequence, the second terms in the large square brackets in (3.8),(3.9),(3.12) and (3.13) are absent. The
 equations from (3.14) onward are correct and all results and conclusions of the paper
  are unchanged. I am indepted to Y.Keilman for pointing out this mistake. 
  \par As shown in the Appendix of Ref.~\cite{JHFASIV}, the formulae (3.1) and (3.2) for `parabolic' 
    acceleration' correspond, not, as hitherto supposed~\cite{Marder,Nikolic}, to a constant proper-frame
    acceleration, $a$, but to one that is proportional to $\gamma = 1/\sqrt{1+(at/c)^2}$. 
   \par The velocity transformation formulae (7.6) and (7.7) are special cases of the `relativistic
     relative velocity transformation relation' (RRVTR) that is derived and discussed, in comparison 
      with the conventional relativistic parallel velocity transformation formula (6.23), in 
     Refs.~ \cite{JHFSTP3,JHFRECP}.

\pagebreak

\end{document}